\documentclass[]{aastex631}

\usepackage{multirow,amsmath,threeparttable}
\usepackage{amssymb}

\newcommand{\pc}{\mathrm{pc}}
\newcommand{\yr}{\mathrm{yr}}

\newcommand{\erg}{\mathrm{erg}}
\newcommand{\g}{\mathrm{g}}
\newcommand{\s}{\mathrm{s}}

\newcommand{\percc}{\mathrm{cm}^{-3}}
\newcommand{\cm}{\mathrm{cm}}
\newcommand{\EM}{\mathrm{EM}}

\newcommand{\iras}{IRAS 00500+6713}

\newcommand{\WD}{WD J005311}
\newcommand{\red}{\textcolor{black}}
\received{XXX}
\revised{YYY}
\accepted{ZZZ}

\submitjournal{ApJ}
\begin{document}

\title{A dynamical model for IRAS 00500+6713: the remnant of a type Iax supernova SN 1181 \\ hosting a double degenerate merger product WD J005311}
\correspondingauthor{Takatoshi Ko}
\email{ko-takatoshi@resceu.s.u-tokyo.ac.jp}

\author[0000-0001-6842-5441]{Takatoshi Ko}
\affiliation{Research Center for the Early Universe, Graduate School of Science, The University of Tokyo, Bunkyo-ku, Tokyo 113-0033, Japanz2}
\affiliation{Department of Astronomy, School of Science, The University of Tokyo, 7-3-1 Hongo, Bunkyo-ku, Tokyo 113-0033, Japan}
\affiliation{Astrophysical Big Bang Laboratory, RIKEN, 2-1 Hirosawa, Wako, Saitama 351-0198, Japan}
\author[0000-0002-8152-6172]{Hiromasa Suzuki}
\affiliation{Department of Physics, Konan University, 8-9-1 Okamoto, Higashinada, Kobe, Hyogo 658-8501, Japan}

\author{Kazumi Kashiyama}
\affiliation{Astronomical Institute, Tohoku University, Sendai 980-8578, Japan}
\affiliation{Kavli Institute for the Physics and Mathematics of the Universe, The University of Tokyo, Kashiwa 277-8583, Japan}

\author{Hiroyuki Uchida}
\affiliation{Department of Physics, Kyoto University, Kitashirakawa Oiwake-cho, Sakyo, Kyoto 606-8502, Japan}

\author{Takaaki Tanaka}
\affiliation{Department of Physics, Konan University, 8-9-1 Okamoto, Higashinada, Kobe, Hyogo 658-8501, Japan}

\author[0000-0002-6347-3089]{Daichi Tsuna}
\affiliation{TAPIR, Mailcode 350-17, California Institute of Technology, Pasadena, CA 91125, USA}
\affiliation{Research Center for the Early Universe, Graduate School of Science, The University of Tokyo, Bunkyo-ku, Tokyo 113-0033, Japan}

\author{Kotaro Fujisawa}

\affiliation{Department of Liberal Arts, Tokyo University of Technology, 5-23-22 Nishikamata, Ota-ku, Tokyo 144-8535, Japan}  
\affiliation{Research Center for the Early Universe, Graduate School of Science, The University of Tokyo, Bunkyo-ku, Tokyo 113-0033, Japan}

\author{Aya Bamba}
\affiliation{Department of Physics, Graduate School of Science, The University of Tokyo, 7-3-1 Hongo, Bunkyo-ku, Tokyo 113-0033, Japan}
\affiliation{Research Center for the Early Universe, Graduate School of Science, The University of Tokyo, Bunkyo-ku, Tokyo 113-0033, Japan}
\affiliation{Trans-Scale Quantum Science Institute, The University of Tokyo, Tokyo  113-0033, Japan}

\author{Toshikazu Shigeyama}
\affiliation{Research Center for the Early Universe, Graduate School of Science, The University of Tokyo, Bunkyo-ku, Tokyo 113-0033, Japan}
\affiliation{Department of Astronomy, School of Science, The University of Tokyo, 7-3-1 Hongo, Bunkyo-ku, Tokyo 113-0033, Japan}

\begin{abstract}
IRAS 00500+6713 is a hypothesized remnant of a type Iax supernova SN 1181. Multi-wavelength observations have revealed its complicated morphology; a dusty infrared ring is sandwiched by the inner and outer X-ray nebulae.  We analyze the archival X-ray data taken by XMM-Newton and Chandra to constrain the {angular radius}, mass, and metal abundance of the X-ray nebulae, and construct a theoretical model describing the dynamical evolution of IRAS 00500+6713, including the effects of the interaction between the SN ejecta and the intense wind enriched with carbon burning ashes from the central white dwarf (WD) J005311. We show that the inner X-ray nebula corresponds to the wind termination shock while the outer X-ray nebula to the shocked interface between the SN ejecta and the interstellar matter. The observed X-ray properties can be explained by our model with an {ejecta kinetic} energy of $E_\mathrm{ej} = (0.77 \mbox{--} 1.1)\times 10^{48}$~erg, an ejecta mass of $M_\mathrm{ej} = 0.18\mbox{--}0.53~M_\odot$, if the currently observed wind from WD J005311 started to blow $t_\mathrm{w} \gtrsim 810$ yr after the explosion, i.e., approximately after A.D. 1990. The inferred SN properties are compatible with those of Type Iax SNe and the timing of the wind launch may correspond to the Kelvin-Helmholtz contraction of the oxygen-neon core of WD J005311 that triggered a surface carbon burning. Our analysis supports that IRAS 00500+6713 is the remnant of SN Iax 1181 produced by a double degenerate merger of oxygen-neon and carbon-oxygen WDs, and WD J005311 is the surviving merger product. 
\end{abstract}


\section{Introduction}\label{sec:intro}
A double white dwarf (WD) merger is an important event because it might lead to a supernova (SN) explosion if the total mass is close to or exceeds the Chandrasekhar limit \citep[e.g.,][]{1984ApJS...54..335I}. Another possibility was proposed \citep{1985ApJ...297..531N,1985A&A...150L..21S} that it results in a gravitational collapse to form a neutron star without a violent explosion such as an SN even when the total mass exceeds the Chandrasekhar limit. We do not know what will happen as a consequence of the merging especially when the total mass is close to or exceeds the Chandrasekhar limit. 

\iras, a Galactic infrared nebula hosting an extremely hot WD J005311 at the center{, which was reported by \cite{2019Natur.569..684G}}, may provide definitive testimony as to the fate of this kind of event. Optical spectroscopic observations identified an intense wind blowing from the WD~\citep{2019Natur.569..684G,Lykou2022}: the wind velocity reaches $v_\mathrm{w} = 15,000\,\mathrm{km\,s^{-1}}$, which is significantly faster than the escape velocity from an ordinary mass WD, and the mass loss rate is as high as those of massive stars, $\dot M_\mathrm{w} \sim 10^{-6}\,M_\odot\,\mathrm{yr^{-1}}$. The chemical composition of the wind is dominated by oxygen, and the neon mass fraction is significantly larger than the solar abundance, indicating an enrichment of carbon-burning ashes. Given the peculiarities of the wind, \WD~has been proposed to be a super- or near-Chandrasckar limit WD with a strong magnetic field and a fast spin, formed via a double degenerate WD merger~\citep{2019Natur.569..684G,2019ApJ...887...39K}. In fact, the photometric properties, i.e., an effective temperature of $T_\mathrm{eff} \sim 20{0},000\,\mathrm{K}$ and a bolometric luminosity of $L_\mathrm{bol}\sim 40,000\,L_\odot$~\citep{2019Natur.569..684G,Lykou2022}, are broadly consistent with a theoretically calculated remnant WD $\sim 1,000\mbox{--}10,000\,\mathrm{yr}$ after the merger~\citep{2016MNRAS.463.3461S,2023MNRAS.524.1031Y,2023ApJ...944L..54W}.

If indeed \WD~is a remnant of a double degenerate WD merger, the surrounding \iras~nebula should hold the information on the merger dynamics and the post-merger evolution of the remnant. In fact, multi-wavelength observations of \iras~have revealed its complicated morphology \citep[e.g.,][]{2019Natur.569..684G,2021ApJ...918L..33R,2023ApJ...945L...4F}. 
{}{From the infrared image taken by the Wide-field Infrared Survey Explorer (WISE) reported by \cite{2019Natur.569..684G}, it can be inferred that a dust-rich ring with an angular radius of $\theta_\mathrm{d} \sim 60\,\mathrm{arcsec}$ is surrounded by a diffuse infrared halo extending to $\theta_\mathrm{out} \sim 140~\mathrm{arcsec}$}. \cite{2021ApJ...918L..33R} conducted a long-slit spectroscopic observation with OSIRIS on board the Gran Telescopio CANARIAS, and measured the expansion velocity of a layer surrounding the dust-rich ring at $\theta_\mathrm{SII} \sim 90\,\mathrm{arcsec}$ as $v_\mathrm{SII} \sim 1,100\,\mathrm{km\,s^{-1}}$ from the {Doppler shift of the [SII] doublet}. Given $\theta_\mathrm{SII}$, $v_\mathrm{SII}$, and the distance to the source $d = 2.3\,\mathrm{kpc}$~\citep{2021AJ....161..147B}, the age of \iras~can be estimated to be $\sim 1,000$ yr. On the other hand, the observation with XM{}{M}-Newton identified the X-ray counterpart of \iras, consisting of the inner nebula enveloped within the dust-rich ring and the outer nebula whose {angular radius} is comparable to the infrared halo~\citep{Oskinova_et_al_20}. The X-ray spectra indicate that both the inner and outer nebulae are also enriched with carbon-burning ashes. 

The outer X-ray nebula likely corresponds to the shocked interface between the interstellar matter~(ISM) and an expanding matter ejected at the merger of the progenitor binary. From the {angular radius} $\theta_\mathrm{out}$ and the emission measure of the outer X-ray nebula $\mathrm{EM}_\mathrm{out}$, \cite{Oskinova_et_al_20} estimated the ejecta mass as $M_\mathrm{ej} \sim 0.1~M_\odot$.
Assuming that the bulk of the nebula is expanding with a velocity comparable to the dust-rich ring, i.e., $\sim 1,000\,\mathrm{km\,s^{-1}}$ \citep[][]{2021ApJ...918L..33R}, the kinetic energy is estimated to be $E_\mathrm{ej} \sim 10^{48}\,\mathrm{erg}$. Such properties are compatible with so-called Type Iax supernovae (SNe Iax; \citealt{2013ApJ...767...57F}), which can be accompanied by the merger of carbon-oxygen and/or oxygen-neon WDs~\citep[see e.g.,][]{2018ApJ...869..140K}. Interestingly, \cite{2021ApJ...918L..33R} and \cite{Lykou2022} pointed out that the sky location and the timing of the hypothesized merger are in accord with the ancient records on a historical Galactic SN, SN 1181. The documented apparent magnitude of SN 1181 is also consistent with SNe Iax, although the uncertainties are large \cite[][]{2023MNRAS.523.3885S}. {}{In addition, recently, \cite{2023ApJ...945L...4F} analyzed deep optical [SII] $\lambda \lambda6716,6713$ images and revealed radially aligned filaments with angular scales of 5 -- 20 arcsec. The authors proposed that these filaments can be formed via the photoionization of the ejecta if the ionizing photons are produced by the dissipation of the wind luminosity and injected into a clumpy region.}

The origin of the inner X-ray nebula is uncertain; the XMM image was reported to be consistent with a point source~\citep{Oskinova_et_al_20}. Given that the intense wind with $v_\mathrm{w} = 15,000\,\mathrm{km\,s^{-1}}$ is blowing \citep[][]{Lykou2022} toward the infrared halo expanding at a slower velocity $v_\mathrm{SII} \sim 1,100\,\mathrm{km\,s^{-1}}$ \citep[][]{2021ApJ...918L..33R}, the wind termination shock would be the most likely possibility.
As such, the intense wind can have significant impacts on the structure of \iras~and accordingly on the values of the physical parameters estimated from the observed morphology. In addition, the history of the wind mass loss is directly connected to the time evolution and the fate of the merger product WD. Thus, it is crucial to construct a dynamical model that cohesively contains the wind, the SN ejecta, the ISM, and intervening multiple shocks 

In this paper, we first analyze the X-ray data on \iras~obtained with XMM-Newton and Chandra to determine the {angular radius}, emission measure, and metal abundance of the X-ray nebulae, taking into account the possible non-collisional ionization equilibrium (non-CIE) effects (Sec. \ref{sec:X-ray}). In particular, we newly obtain a constraint on the {angular radius} of the inner X-ray nebula from the Chandra data. Next, we construct a dynamical model based on the hypothesis that \iras~is the remnant of SN 1181 (Sec. \ref{sec:model}). We consistently solve the dynamics of the inner and outer X-ray nebulae, allowing that the currently observed wind started to blow a finite time after SN 1181. We use the dynamical model to constrain the relevant physical parameters of the system, i.e., the {kinetic} energy and mass of the SN ejecta, the timing of the launch of the wind, from the currently observed multi-wavelength morphology. We discuss the implications of the results of our dynamical model in Sec. \ref{sec:discussion}. 

\section{X-ray data analysis} \label{sec:X-ray}
\subsection{Observations and data reduction}
IRAS~00500+6713 was observed with XMM-Newton in 2019 and 2021 and with Chandra in 2021.
We use \red{all the available archival} data obtained with the CCDs onboard both observatories, XMM EPIC (MOS1, MOS2, and pn; \citealt{turner01, struder01, jansen01}) and Chandra ACIS \citep{garmire97} for the imaging analysis.
For the spectroscopy, we simultaneously use imaging spectroscopic data obtained with XMM EPIC and grating data obtained with XMM RGS \citep{denherder01}. Table~\ref{tab:obs} lists the XMM and Chandra observation logs.
We use the XMM Science Analysis Software (SAS; v19.1.0; \citealt{gabriel04}) package, which also includes the Extended Source Analysis Software (ESAS; \citealt{snowden04}). We process the EPIC data with the SAS tasks \texttt{pn-filter}, \texttt{epchain}, \texttt{emchain}, and \texttt{mos-filter}. The RGS data are processed using the standard pipeline tool \texttt{rgsproc} in the SAS package. We exclude one observation (OBSID: 0872590301) because it shows high count rates over the whole field of view and is affected by background flares. The total effective exposures for the EPIC and RGS are \red{$\approx 76$~ks and $\approx 102$~ks, respectively}. 
For the Chandra data, we process the data with the CIAO (v4.15; \citealt{fruscione06}) tool \texttt{chandra\_repro}. The total effective exposure is $\approx 144$~ks. \red{Considering relatively low statistics of the Chandra data, we use them only for the imaging analysis (Section~\ref{sec::image}). All the five observations are merged for use in the analysis.}

In the data analysis, we use the HEASoft (v6.20; \citealt{heasarc14}), XSPEC (v12.9.1; \citealt{arnaud96}), and AtomDB (v3.0.9; \citealt{smith01, 2012ApJ...756..128F}). The {\it cstat} \citep{cash79} and {\it wstat}\footnote{\red{Please refer to the XSPEC documentation (\url{https://heasarc.gsfc.nasa.gov/xanadu/xspec/manual/node319.html}).}} implemented in XSPEC are used in our spectral studies. In this section, uncertainties in the text, figures, and tables indicate 1$\sigma$ confidence intervals. The upper \red{and lower} limits presented in this section are at a 95\% confidence level.

\begin{table}[htb]
    \centering
    \caption{\red{XMM and Chandra observation logs of IRAS~00500+6713}}
    \label{tab:obs}
    \begin{tabular}{c c c c c c}
\hline\hline
 & OBSID & Date &  \multicolumn{2}{c}{Exposure (ks)}\tablenotemark{a} & \red{PI name} \\
 &  &  & \red{MOS2} & \red{RGS1} & \\\hline
XMM & 0841640101 & 2019 Jul 8 & 14.2 & 19.5 & L. Oskinova \\
& 0841640201 & 2019 Jul 24 & 13.4 & 16.7 & L. Oskinova \\
& 0872590101 & 2021 Jan 8 & 6.8 & 11.8 & L. Oskinova\\
& 0872590201 & 2021 Jan 10 & 10.9 & 12.9 & L. Oskinova \\
& 0872590301\tablenotemark{b} & 2021 Jan 14 & 4.0 & 7.9 & L. Oskinova \\
& 0872590401 & 2021 Jan 16 & 7.7 & 17.2 & L. Oskinova \\
& 0872590501 & 2021 Jan 20 & 10.4 & 12.7 & L. Oskinova\\
& 0872590601 & 2021 Jan 18 & 12.2 & 14.6 & L. Oskinova \\\hline
Chandra & 23419 & 2021 May 12 & \multicolumn{2}{c}{27.7} & L. Oskinova \\
& 24342 & 2021 May 17 & \multicolumn{2}{c}{14.8} & L. Oskinova \\
& 24343 & 2021 May 14 & \multicolumn{2}{c}{27.7} & L. Oskinova \\
& 24344 & 2021 Oct 21 & \multicolumn{2}{c}{38.0} & L. Oskinova \\
& 24345 & 2021 Dec 15 & \multicolumn{2}{c}{20.8} & L. Oskinova \\
& 25045 & 2021 May 17 & \multicolumn{2}{c}{14.8} & L. Oskinova \\\hline
\end{tabular}
\tablenotetext{a}{\red{Effective exposures of MOS2 and RGS1 are presented representing the instruments.}}
\tablenotetext{b}{\red{Excluded from the analysis due to a high background rate.}}
\end{table}


\subsection{Analysis and results}
\subsubsection{Spatial distributions}\label{sec::image}
\begin{figure}[ht]
    \centering
    \includegraphics[width=16cm]{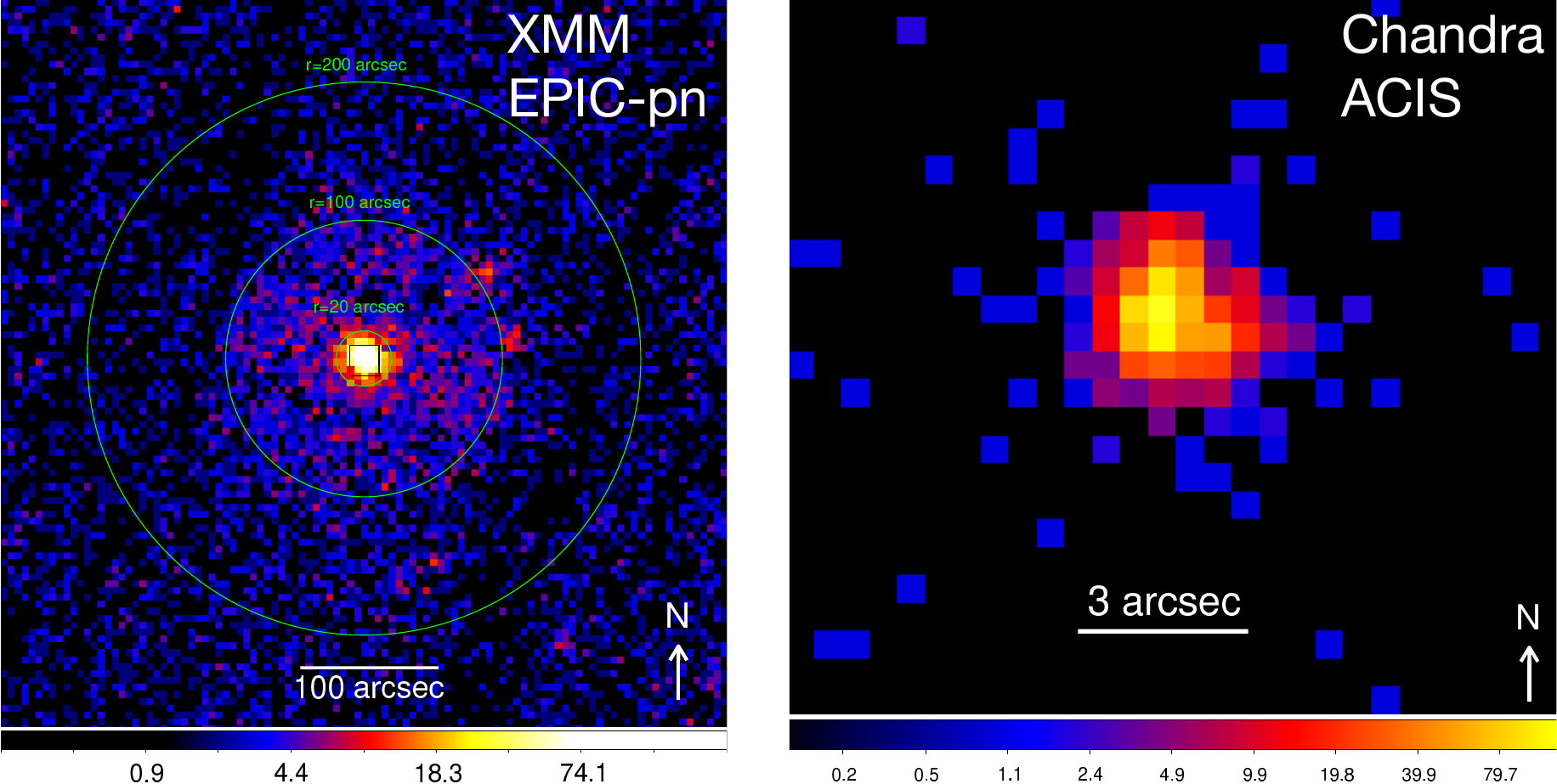}
    \caption{0.5--5.0~keV images of IRAS~00500+6713 obtained with XMM EPIC-pn ({\it left}) and Chandra ACIS ({\it right}). For the XMM image, all the observations but one with OBSID 0872590301 are merged. All the observations are used to make the Chandra image. In the left panel, circles with radii of 20, 100, and 200 arcsec are overlaid in green. The former two correspond to the approximate spatial extents of the central and diffuse sources. \red{The central black square in the XMM image shows the range of the Chandra image}. The images are shown on a logarithmic scale. }
    \label{fig:image}
\end{figure}

\begin{figure}[ht]
    \centering
    \includegraphics[width=16cm]{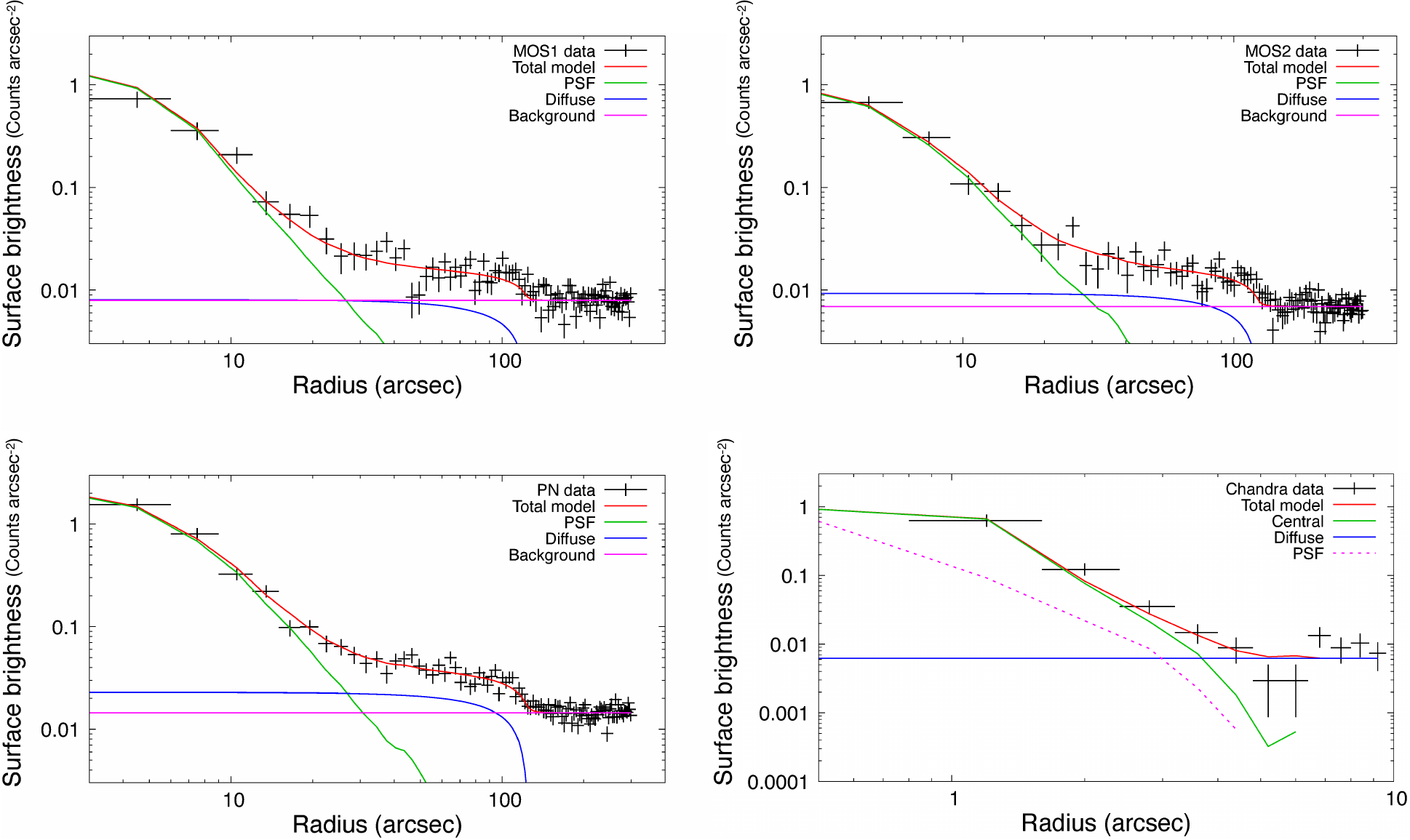}
    \caption{Radial flux profiles of IRAS~00500+6713 obtained with XMM MOS1, MOS2, pn, and Chandra ACIS. The XMM data are extracted from one observation in 2019 (OBSID: 0841640101). All the observations are merged for the Chandra data. The energy range for extraction is 0.5--5.0~keV. The solid lines represent the best-fit model components.}
    \label{fig:prof}
\end{figure}

Figure~\ref{fig:image} shows 0.5--5.0~keV X-ray images obtained with XMM EPIC-pn and Chandra ACIS. One can see a central source \red{($\lesssim 20\arcsec$)} and a faint diffuse emission around it \red{($\lesssim 100\arcsec$)}. \red{We exclude data in the lower energies considering less accurate detector calibrations, and in the higher energies because of higher background rates.} The central source is found to be slightly extended with the spatial resolution of Chandra ($\approx 0\farcs5$). \red{The diffuse emission extending to $\sim 100\arcsec$ is not visible with Chandra because of the high detector background rate, which is $\approx 10$ times higher than the expected diffuse flux.}
In order to evaluate the spatial distribution of the source, we extract radial profiles around the central source. The central position for extraction is set to (R.A., Decl.$)_{2000}=(13\fdg2967, \, 67\fdg5007)$ by referring to the SIMBAD catalog\footnote{\url{http://simbad.u-strasbg.fr/simbad/sim-id?Ident=IRAS+00500\%2B6713}}. We take into account spatially-dependent exposures \red{(depending on off-axis angles)}, bad pixels, and gaps between the CCDs. For XMM EPIC, we obtain radial flux profiles for each instrument for each observation. For Chandra ACIS, we merge all the observations and extract radial profiles considering almost the same pointing directions and observation dates close to each other. In Figure~\ref{fig:prof}, we show flux profiles extracted from one XMM observation in 2019 (OBSID: 0841640101) and the merged Chandra data.

We compare the observed distributions with a phenomenological model. The instrumental point spread functions (PSFs), i.e., the radial profiles of an ideal point source on the CCDs, are calculated for the source position with the SAS tool {\tt psfgen} and CIAO tool \texttt{simulate\_psf}. We assume a simple emission model for the central and diffuse sources: a sphere with a uniform and isotropic emissivity over the whole volume.
The model for the radial flux profile ($F(r)$, where $r$ is the radius) is described as
\begin{equation}
    F(r) = P(r, \theta_{\rm in}) + D(r, \theta_{\rm out}) + B,
\end{equation}
with $P(r, \theta_{\rm in})$, $D(r, \theta_{\rm out})$, and $B$ is the fluxes of the central source, diffuse source, and background (assumed to be uniform), respectively. The parameters $\theta_{\rm in}$ and $\theta_{\rm out}$ indicate the spatial extents of the central and diffuse sources, respectively. $P(r, \theta_{\rm in})$ and $D(r, \theta_{\rm out})$ are spherical emission convolved with the instrumental PSFs.
We mainly focus on evaluation of $\theta_{\rm in}$ and $\theta_{\rm out}$.

We use the MCMC (Markov Chain Monte Carlo) algorithm\footnote{The C++ library MCMClib (\url{https://github.com/kThohr/mcmc}) is partly used.} to determine the best-fit model parameters and their confidence ranges. To perform a precise evaluation, the profiles extracted from all the instruments in all the observations (21 in total) are modeled simultaneously via linked $\theta_{\rm in}$ and $\theta_{\rm out}$ for the XMM data. The normalizations of $P(r, \theta_{\rm in})$ and $D(r, \theta_{\rm out})$, and $B$ for individual instruments are treated as free parameters. Radial ranges of 0--300 and 0--10 arcsec are used for the modeling of the XMM and Chandra data, respectively.
As a result, we obtain $\theta_{\rm in} < 1.5$ arcsec and $\theta_{\rm out} = 131^{+2}_{-2 -2}$ arcsec\footnote{Value $\pm$ (statistic error) $-$ (pointing accuracy) is presented. The reference for the pointing accuracy is XMM Users' Handbook (\url{https://www.mssl.ucl.ac.uk/www_xmm/ukos/onlines/uhb/XMM_UHB/node104.html}); \citealt{vries15}.} with the XMM data.
\red{We obtain the corresponding $-2 \log L/\text{d.o.f.} = 2597 / 2026$, where $L$ and d.o.f. are the likelihood value and degree of freedom, respectively.
This likelihood value is marginally acceptable considering that we are using a simple phenomenological model.}
\red{The central source is consistent with a point-like source for XMM.}
The spatial extent of the diffuse source is similar to that in infrared \citep{2019Natur.569..684G} as pointed out by \cite{Oskinova_et_al_20}. On the other hand, we are able to constrain the radius \red{of the central source} as $\theta_{\rm in} = 1.52^{+0.08}_{-0.08 -0.67}$ arcsec\footnote{The reference for the pointing accuracy is \url{https://cxc.harvard.edu/cal/ASPECT/celmon/}.} with Chandra. \red{The corresponding $-2 \log L/\text{d.o.f.} = 29.0 / 9$, is also marginally acceptable.}
The best-fit models are overlaid in Figure~\ref{fig:prof}.
We also search for a possible time variation of the spatial distributions by separately modeling the XMM flux profiles in 2019 and 2021. We find results consistent with each other, $\theta_{\rm in} < 1.5$ arcsec and $\theta_{\rm out} = 132 \pm 2$ arcsec in 2019 and $\theta_{\rm in} < 1.5$ arcsec and $\theta_{\rm out} = 129 \pm 3$ arcsec in 2021{}{, which means that there is almost no time variation in the sizes of the X-ray emitting regions. Throughout this paper, we adopt the value $\theta_{\rm out} = 131^{+2}_{-2 -2}$ arcsec obtained using all the observations simultaneously.}

The background rates $B$ determined from the XMM data, e.g., $\approx 0.0023$ counts~s$^{-1}$~arcmin$^{-2}$ (0.5–5.0 keV) for MOS1 in 2021, are consistent with or slightly higher than the pure particle background rate \citep{kuntz08}. This is reasonable considering the contribution of the sky background. The diffuse plus background flux $D(r,\theta_{\rm out}) + B$ for the Chandra data are also consistent with the sum of the diffuse flux determined from the XMM data and the particle background rate \citep{bartalucci14, suzuki21b}. {}{Using the observed constraints on $\theta_{\rm in} = 1.52^{+0.08}_{-0.08 -0.67}$ arcsec with 1 sigma uncertainties, the minimum and maximum values of $\theta_{\rm in}$ are 0.77 arcsec and 1.60 arcsec, respectively. Adopting the  Gaia distance of 2.3 kpc, the minimum and maximum physical radii are 0.0087~pc and 0.018~pc. These values are used to compare with our model. The spread of the inner source $r_{\rm in}=0.0087$--$0.018~\pc$ is much larger than the photospheric radius $\sim 1\times 10^{10}$ cm \citep{2019Natur.569..684G}. This indicates that the inner shocked region has certainly spreaded, likely due to the wind from the central WD. However, since the wind-crossing time $r_{\rm in}/v_{\rm w} \lesssim 1 $yr is much shorter than the SN age, this region may have been formed very recently.}

\subsubsection{Spectral properties of the thermal plasmas}\label{sec]spectral_prop}
We here extract spectra and estimate the temperatures, metal abundances, ionization time scale, and emission measures of the central and diffuse sources. We use the XMM data, which have \red{more than twice} greater statistics than the Chandra data. Spectral extraction regions for the EPIC data are selected based on the imaging analysis results (Figure~\ref{fig:prof}) and are shown in Figure~\ref{fig:image}. A circle with a 20 arcsec radius is used for the central source, and an annulus with inner and outer radii of 30 and 100 arcsec, respectively, is assumed for the diffuse source. As the background, we use the diffuse source and an annulus region with inner and outer radii of \red{140} and 200 arcsec, respectively.
\red{The contamination of the diffuse emission to the background region is $< 1\%$ of the total flux according to the results of our imaging analysis.}
The central position for extraction is the same as that used for the extraction of the radial profiles.
Since the RGS spectrum has much poorer statistics \red{($<1\%$ of the EPIC data)}, we only use it as a consistency check for the spectral analysis of the bright central source.
We merge all the spectra in each year and for each instrument: the 1st- and 2nd-order spectra are also combined for the RGS data.
The spatial variation of the detector background is negligible considering the on-axis position and sizes of our analysis regions \citep{kuntz08}.

As the spectral model, we consider \red{optically-thin} thermal plasmas in two different assumptions of metalicity by referring to \cite{Oskinova_et_al_20}: near solar and pure metal. Under each assumption, \red{we try several models with different free parameters and finally adopt the simplest} model that explains the data sufficiently well. \red{Note that we do not pre-assume the abundance patterns reported in other wavelengths.}
\red{We find an absorbed ionizing plasma model (${\tt tbabs} \times {\tt nei}$ in XSPEC) explain both central and diffuse sources}.
\red{The treatment of the model parameters for the central source is as follows:}
in the near-solar case, the hydrogen column density ($N_{\rm H}$), electron temperature ($k_\mathrm{B}T_{\rm e}$), ionization timescale ($\tau$), emission measure, and abundances of O, Mg, and Fe ($=$Ni) are treated as free parameters. The other metal abundances are fixed to solar. {}{Here $k_\mathrm{B}$ denotes the Boltzmann constant.}
In the pure-metal case, the same model but without H, He, N, Ar, and Ca is used.\footnote{In XSPEC, we set the abundance of C to 2800 to suppress the contribution of the finite continuum emission of H and He, which cannot be removed in XSPEC. \red{Note that this artificial scaling is canceled in calculations of emission measures and metal abundances.}}
\red{For the diffuse source}, the free parameters are the same as above but the abundance of O is fixed to solar and Ne is set free.
The spectra obtained with all the instruments in 2019 and 2021 (six in total) are modeled simultaneously. We use the energy ranges of 0.5--5.0~keV and 0.5--2.0~keV for the spectral fit of the central and diffuse sources, respectively, considering their statistics.

\begin{figure}[htb]
    \centering
    \includegraphics[width=16cm]{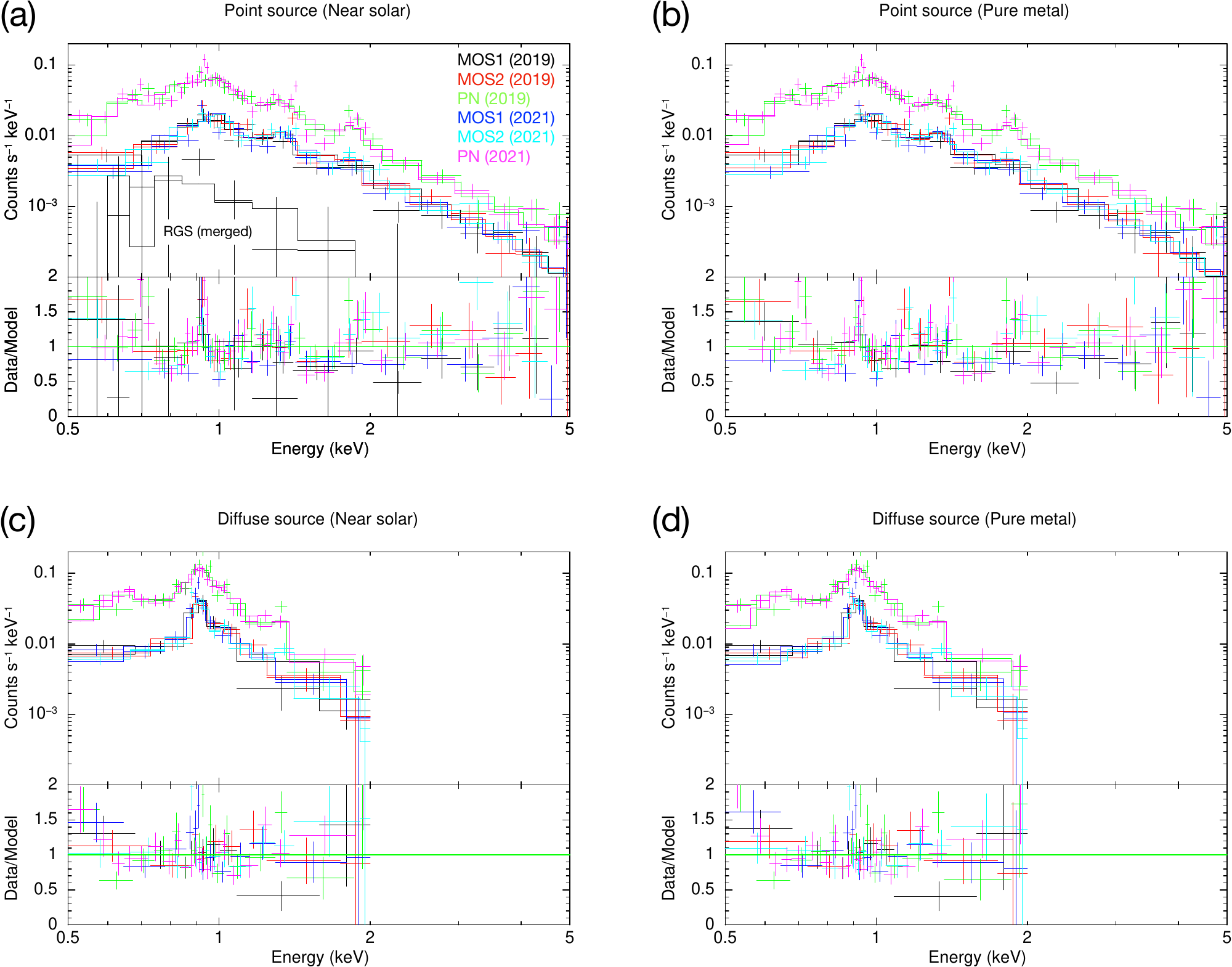}
    \caption{Energy spectra extracted from the central point-like source (panels (a) and (b)) and surrounding diffuse source (panels (c) and (d)). Multiple observations in each year are merged. The solid lines show the best-fit models in the two cases (near solar and pure metal).
    The background spectra are subtracted from all the data. The XMM RGS spectrum, which is used for a consistency check, is presented in the panel (a).}
    \label{fig:spec}
\end{figure}

The spectra and best-fit models are presented in Figure~\ref{fig:spec}. Table~\ref{tab:specb} shows the best-fit parameters. Both models explain well both central and diffuse sources \red{with acceptable fit statistics.}
\red{The emission measures (EMs; hereafter $\EM_{\rm in}$ and $\EM_{\rm out}$ for the central and diffuse sources, respectively) derived for the pure-metal and near-solar plasmas differ by two orders of magnitude because of the difference in the total ion emissivity.}
\red{The abundance ratios are C/O=$0.014\pm0.003$, Ne/O=$0.005\pm0.001$, Mg/O=$0.0010\pm0.0004$, Fe/O=$0.005\pm0.002$ for the central source, and C/O=0.43 (fixed), Ne/O=$0.41\pm0.07$, Mg/O=$0.04\pm0.01$, Fe/O$< 0.003$ for the diffuse source (all in number). The mass fractions in the pure-metal case are presented in Table~\ref{tab:mass}.}
\cite{Oskinova_et_al_20} applied a two-temperature, collisional-ionization-equilibrium thermal plasma model to these sources. \red{Their assumption of the abundance pattern is partly based on the infrared observations \citep{2019Natur.569..684G} and thus different from ours. Although the mass fractions in this work differ from previous studies \citep{2019Natur.569..684G, Oskinova_et_al_20, Lykou2022}, the tendency is similar: C/O$< 1$ and Ne/O$< 1$ for the central source and C/O$\sim 1$ and Ne/O$\sim 1$ for the diffuse source.}
Compared to \cite{Oskinova_et_al_20}, the electron temperatures we obtain here lie between their higher and lower temperatures. The absorption column densities are similar.

\red{Note that we use the data with approximately twice better statistics than those used in \cite{Oskinova_et_al_20}. \cite{Oskinova_et_al_20} suggested that an additional non-thermal component might contribute to the central region. If we add an additional non-thermal component described with a power-law with the fixed index of 2.4 following \cite{Oskinova_et_al_20}, we indeed obtain an improved fit with $C$-statistic/d.o.f. = 4294/5408. However, since the fit statistics before adding a non-thermal component are already acceptable, we cannot claim that the central source requires a non-thermal component.}
We also search for possible spectral changes as a function of time but find no such evidence.

\begin{deluxetable*}{c c c c c c c c c c c}[htb]
    \tablecaption{Best-fit spectral parameters for the point-like and diffuse sources \label{tab:specb}
}
\tablewidth{16cm}
\tabletypesize{\scriptsize}
\tablehead{
 \colhead{} & \colhead{} & \colhead{$N_{\rm H}$($10^{22}\,{\rm cm}^{-2}$)} & \colhead{$k_\mathrm{B}T_{\rm e}$(keV)} & \colhead{\red{$\tau (10^{10} \text{s cm}^{-3})$\tablenotemark{a}}} & \colhead{O\tablenotemark{b}} & \colhead{Ne\tablenotemark{b}} & \colhead{Mg\tablenotemark{b}} & \colhead{Fe ($=$Ni)\tablenotemark{b} } & \colhead{\red{$\mathrm{EM (10^{52} \text{cm}^{-3})}$\tablenotemark{c}}} & \colhead{$C$-stat/d.o.f.}\tablenotemark{d}} 
\startdata
 Point source & Near solar & 0.78$\pm$0.05 & 1.30$\pm$0.08 & 7.9$\pm$1.3 & 31$\pm$8 & 1 (fixed) & 0.68$\pm$0.25 & 2.81$\pm$0.63  & (1.8$\pm$0.4)$\times10^{2}$ & 4406/5409\\
 & Pure metal  & 0.75$\pm$0.05 & 1.21$\pm$0.06 & 9.5$\pm$1.4 & 37$\pm$9 & 1 (fixed) & 0.70$\pm$0.25 & 2.62$\pm$0.62 & 6.3$\pm$1.1  & 4405/5409 \\
 Diffuse source & Near solar & 0.51$\pm$0.05 & 0.56$\pm$0.12 & $5.9\pm4.0$ & 1 (fixed) & 2.67$\pm$0.45 & 1.04$\pm$0.27 & 0.04$\pm$0.04 &  (3.7$\pm$1.0)$\times10^{2}$ & 2113/1809 \\
 & Pure metal & 0.53$\pm$0.05 & 0.29$\pm$0.03 & $>53$ & 1 (fixed) & 1.57$\pm$0.27 & 0.74$\pm$0.24 & 0.04$\pm$0.04 &  3.6$\pm$0.8 & 2105/1809 \\
\enddata
\tablenotetext{a}{Ionization timescale.}
\tablenotetext{b}{Metal abundance normalized by that of C, i.e., (O/O$_{\odot}$)/(C/C$_{\odot}$).}
\tablenotetext{c}{Emission measure, $n_{\rm e} n_{\rm ion} V$. The parameters $n_{\rm e}$, $n_{\rm ion}$, and $V$ represent electron and ion number densities, and plasma volume, respectively. A distance of 2.3 kpc is assumed.}
\tablenotetext{d}{\red{$C$-stat and d.o.f. represent the $C$-statistic value and degree of freedom, respectively.}}
\end{deluxetable*}

\begin{deluxetable}{c c c }[htb]
    \tablecaption{\red{Mass fractions of the central and diffuse sources derived in the pure-metal case} \label{tab:mass}}
    \tablehead{
    \colhead{Element} & \colhead{Central} & \colhead{Diffuse} } 
    \startdata
    C & 0.008$^{+0.072}_{-0.008}$ & 0.18 \\
    O & $0.97^{+0.03}_{-0.24}$ & 0.56 \\
    Ne & $0.005^{+0.045}_{-0.005}$ & $0.16\pm0.03$ \\
    Mg & $0.004^{+0.037}_{-0.004}$ & $0.09\pm0.02$ \\
    Fe & $0.010^{+0.092}_{-0.010}$ & $0.004\pm0.004$ \\
    \enddata
\end{deluxetable}

\begin{table}[]
    \centering
        \caption{Parameters in our dynamical model for IRAS 00500+671{3. The references to the parameters in the upper table are as follows: a \citep{Gaia_eDR3}; b \citep{Lykou2022}; c (This work); d {}{(WISE image reported by \citealt{2019Natur.569..684G})}; e \citep{2021ApJ...918L..33R}}.}
    \begin{tabular}{c c c}
    \hline
    & {parameters estimated from observations} &  \\
    \hline 
     distance & $d$ [kpc]$^{*\rm a}$ &$2.3^{+0.1}_{-0.1}$  \\
     wind mass loss rate& $\dot M_\mathrm{w}$ [$M_\odot\,\mathrm{yr^{-1}}$]$^{*\rm b}$  &{$7.5\times 10^{-7} \mbox{--}  4\times 10^{-6}$}\\
     wind mechanical luminosity & $L_\mathrm{w}$ [${\rm erg\, s^{-1}}$]$^{*\rm b}$ &$(0.53\mbox{--}2.8)\times 10^{38}$  \\
     \multirow{3}{5.5em}{{angular radius}} & $\theta_{\rm in}$ [arcsec]$^{*\rm c}$ &$0.77-1.6$  \\
     & $\theta_{\rm out}$ [arcsec]$^{*\rm c}$ &{}{$131^{+2}_{-2-2}$}   \\
     & $\theta_{\rm d}$ [arcsec]$^{*\rm d}$ &$60^{+12}_{-12}$  \\
     expansion velocity & $v_\mathrm{SII}$ [${\rm km\,s^{-1}}$]$^{*\rm e}$ &$1100^{+100}_{-100}$  \\
     \multirow{4}{9em}{emission measure} & \multirow{2}{*}{${\rm EM}_{\rm in}$ [$10^{52}\,{\rm cm^{-3}}$]$^{*\rm c}$ }&180$\pm$40~(near-solar)\\ 
     &&6.3$\pm$1.1~(pure metal)\\
     &\multirow{2}{*}{${\rm EM}_{\rm out}$ [$10^{52}\,{\rm cm^{-3}}$]$^{*\rm c}$ } &370$\pm$100~(near-solar)\\
     &&3.6$\pm$0.8~(pure metal)\\
    \hline
    & {parameters estimated from our model} &  \\
    \hline 
    \multirow{2}{*}{power-law index of the SN ejecta density profile} &$n$&6 (fixed)\\
    &$\delta$ &1.5 (fixed) \\
    number density of ISM& $n_\mathrm{ISM}$ [cm$^{-3}$] &$0.097\mbox{--}0.13$  \\
    {ejecta kinetic} energy & $E_\mathrm{ej}$ [$10^{48}$~erg] &$0.77\mbox{--}1.1$  \\
    ejecta mass & $M_\mathrm{ej}$ [$M_\odot$] &$0.18\mbox{--}0.53$  \\
    wind launching time & $t_\mathrm{w}$ [yr] &$810\mbox{--}828$  \\
    
    \hline

    \end{tabular}
    \label{tab:param}
\end{table}

\section{Dynamical Model} \label{sec:model}

\begin{figure}[ht]
    \centering
    \includegraphics[width=1\columnwidth]{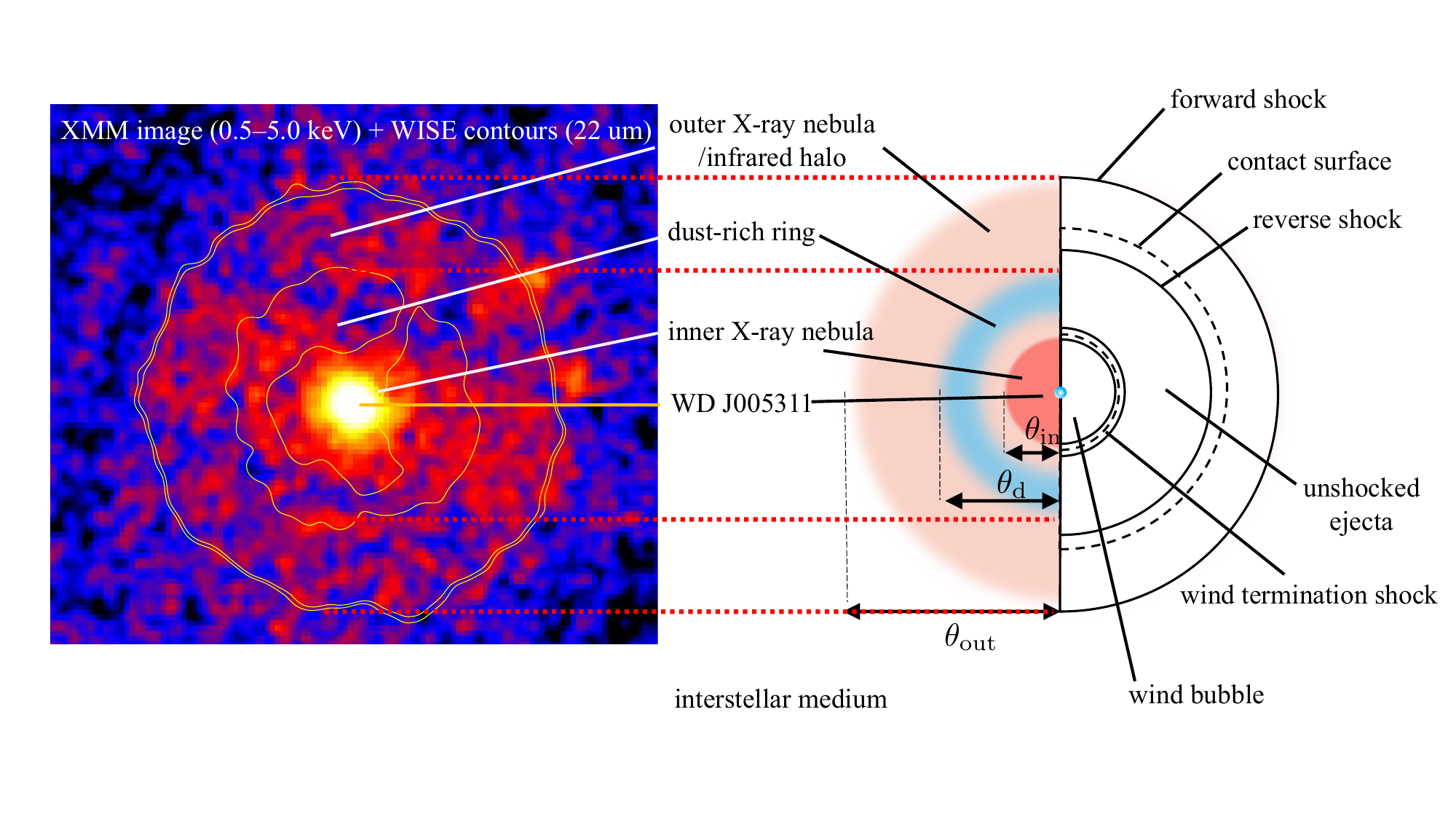}
    \caption{{}{A comparison between an observed image (left panel; X-ray image (XMM) and infrared contours (WISE)) and our schematic picture of \iras~(right panel). The schematic picture includes the morphology of \iras~inferred from multi-wavelength observations (the left semicircle) and our dynamical model (the right semicircle)}. A dusty infrared ring is sandwiched by the inner and outer X-ray nebulae. We assume that the inner X-ray nebula corresponds to the wind termination shock while the outer X-ray nebula to the shocked interface between the ejecta of SN 1181 and the interstellar matter, and the dust ring corresponds to the unshocked SN ejecta.}
    \label{fig:model}
\end{figure}

In the previous section, we have determined the amounts and spatial extents of the X-ray emitting plasma in the inner and outer nebulae of \iras, i.e., the emission measures ($\mathrm{EM}_\mathrm{in}$ and $\mathrm{EM}_\mathrm{out}$) and the {angular radii} ($\theta_\mathrm{in}$ and $\theta_\mathrm{out}$). 
Here we construct a dynamical model of \iras~for interpreting the currently observed X-ray properties {assuming that this nebula is the remnant of SN 1181}. 
{}{The schematic picture of the overall structure is shown in the right part of  Figure \ref{fig:model}.} The inner shocked region is produced by the interaction between the ejecta of SN 1181, set to be launched at $t = 0$, and the wind from the central WD that started blowing at $t = t_\mathrm{w}$. On the other hand, the outer shocked region is produced by the interaction between the SN ejecta and the surrounding ISM. {A comparison between this schematic picture and the observed image is shown on the left of Figure \ref{fig:model}.} We follow the dynamical evolution of the inner and outer X-ray nebulae from the onset of the SN explosion ($t = 0$) through $t = 845\,\mathrm{yr}$ ($t=840$ yr corresponds to A.D. 2021) to determine the characteristic quantities of the system, i.e., the {ejecta kinetic} energy $E_\mathrm{ej}$ and the ejecta mass $M_\mathrm{ej}$ of SN 1181 and the timing $t_\mathrm{w}$ of the launch of the wind that reproduce the X-ray properties. 

We note that other than EMs and $\theta$s, there are observed properties of \iras ~that might be useful to constrain the model. For example, the electron temperatures $k_\mathrm{B}T_\mathrm{e}$ and the ionization timescales $\tau$ have been estimated. 
In addition, the previous mid-infrared observations determined the angular radius $\theta_\mathrm{d}$ of the dust-rich ring~\citep{2019Natur.569..684G}, and the {expansion velocity of the shock-excited ejecta $v_\mathrm{SII}$}~\citep{2021ApJ...918L..33R}. We mainly use these pieces of information to verify the model that fits the EMs and $\theta$s of the X-ray nebulae. The ranges of physical parameters that best reproduce the X-ray observations are listed in Table \ref{tab:param}.

\subsection{Density profiles}\label{sec:setup}
In order to calculate the evolution of this system, the density profile of each region should be given. The density profile of the SN ejecta {is} often described by a double power law~\citep[][]{1989ApJ...341..867C,1999ApJ...510..379M,2013MNRAS.435.1520M}; 
\begin{equation}\label{eqn:rho_ej1}
    \rho_\mathrm{ej}(t,\ r) = \rho_\mathrm{ej,br}(t) \times
    \begin{cases}
        (r/r_\mathrm{br})^{-\delta} & r < r_\mathrm{br}, \\
        (r/r_\mathrm{br})^{-n} & r > r_\mathrm{br},
    \end{cases}
\end{equation}
where {$r_{\rm br}$ is the radius where the exponent changes, and $\rho_\mathrm{ej,br}$ is the density at $r = r_{\rm br}$. These parameters are obtained from the total mass ($M_{\rm ej}$) and kinetic energy ($E_{\rm ej}$) of the SN ejecta as follows \citep[][]{1989ApJ...341..867C,1999ApJ...510..379M,2013MNRAS.435.1520M}:}
\begin{equation}
\rho_\mathrm{ej,br}(t) = t^{-3}(r_\mathrm{br}/tg)^{-n},
\end{equation}
\begin{equation}\label{eqn:r_star}
    r_{\rm br} = \left[\frac{2(5-\delta)(n-5)E_{\rm ej}}{(3-\delta)(n-3)M_{\rm ej}}\right]^{\frac{1}{2}}t,
\end{equation}
\begin{equation}\label{eqn:g^n}
    g^{n} =  \frac{1}{4\pi(n-\delta)}\frac{\left[2(5-\delta)(n-5)E_{\rm ej}\right]^{\frac{n-3}{2}}}{\left[(3-\delta)(n-3)M_{\rm ej}\right]^{\frac{n-5}{2}}}.
\end{equation}
{Here $g$ is a constant obtained for a given $\delta$, $n$, $E_\mathrm{ej}$ and $M_\mathrm{ej}$.} {}{In order to have a finite total mass and energy, $n > 5$ and $\delta < 3$ are required \citep[][]{1982ApJ...258..790C}}. 
We assume that the SN ejecta {are} surrounded by ISM with a constant mass density of 
\begin{equation}
    \rho_{\rm ISM} =\mu_{\rm ISM}m_{\rm u}n_{\rm ISM},
\end{equation}\label{eqn:rho_ISM}
where 
\begin{equation}\label{eqn:mu_ISM}
    {\mu_{\rm ISM} = 0.615}
\end{equation}
is the mean molecular weight assuming the solar abundance, $m_{\rm u}$ is the atomic mass unit, and $n_{\rm ISM}$ is the number density of the ISM. 
On the other hand, WD J005311 resides inside the SN ejecta and blows an intense wind with a mass loss rate of $\dot{M}_\mathrm{w}= (7.5\mbox{--}40)\times10^{-7}~M_\odot\,\yr^{-1}$ and velocity of $v_\mathrm{w} = 15,000~\mathrm{km\,s^{-1}}$~\citep{Lykou2022}.
Accordingly, the mechanical luminosity of the wind can be estimated as $L_\mathrm{w} = \dot{M}_{\rm w}v_{\rm w}^2/2 = (0.53\mbox{--}2.8)\times 10^{38}\,\mathrm{erg\,s^{-1}}$.
We assume that the currently observed wind from \WD~started blowing at $t = t_\mathrm{w}$ with a constant mass loss rate. 
The density profile in the wind region is described as
\begin{equation}
    \rho_\mathrm{w} = \frac{\dot M_\mathrm{w}}{4\pi r^2 v_\mathrm{w}}.
\end{equation}

\begin{figure}[ht]
    \centering
    \includegraphics[width=.8\columnwidth]{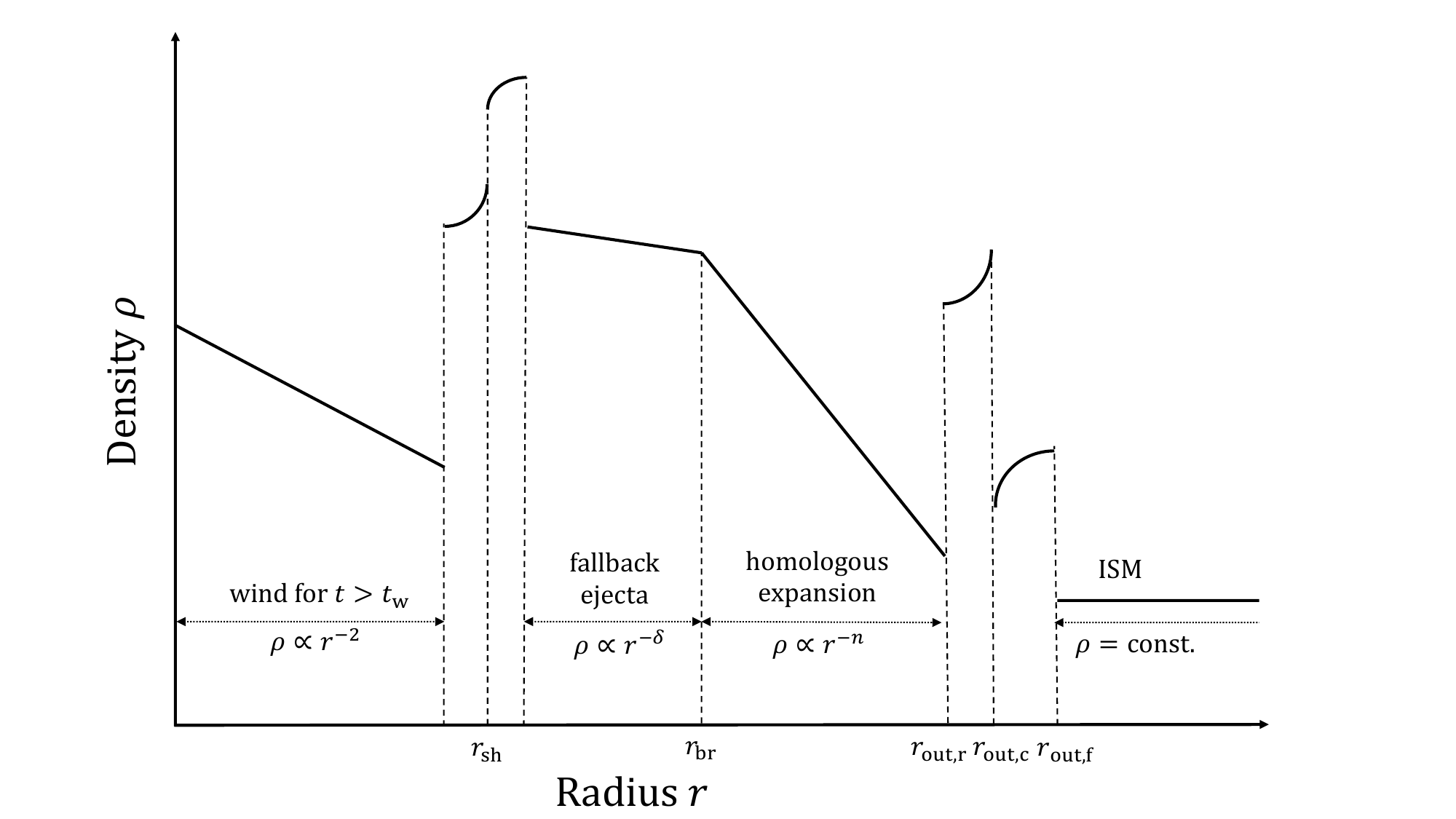}
    \caption{A schematic picture of the radial density profile of IRAS 00500+6713. {The curved regions represent shocked regions and the dotted lines surrounding them represent the forward or reverse shocks.}
    }
    \label{fig:ejecta_density}
\end{figure}
The radial density profile is shown schematically in Figure \ref{fig:ejecta_density}. While the SN ejecta expand in the ISM, the {stellar} wind catches up and collides with the SN ejecta. Two shocked regions form; {the outer one between the SN ejecta and the ISM, and the inner one between the wind and the SN ejecta. The outer and the inner X-ray nebulae are powered by these two shocked regions, which are analyzed in Sections \ref{sec:outer} and \ref{sec:inner}, respectively.} {Even though the near-surface temperature of WD J005311 is expected to be as high as $\gtrsim 10^7\,\mathrm{K}$~\citep{2019ApJ...887...39K}, the observed effective temperature is $T_\mathrm{eff} = 211,000^{+40,000}_{-23.000}\,\mathrm{K}$~\citep{2019Natur.569..684G}. This is due to significant adiabatic cooling in the optically thick wind, and thus the emission from the photosphere cannot explain the observed X-rays from the inner nebula.} In between the two shocked regions, there can be an unshocked layer. In our view, both the dusty infrared ring detected in the WISE image and the [SII] line emission region detected by OSIRIS correspond to this unshocked layer; when the shocks sweep through this layer, dust grains are expected to be destroyed.

\subsection{Outer X-ray nebula}\label{sec:outer}

Let us first consider the outer X-ray nebula. For homologously expanding SN ejecta sweeping a uniform ISM described in Sec. \ref{sec:setup}, we can employ the self-similar solution of \cite{1982ApJ...258..790C} to describe the evolution of the shocked region. 
The actual values of the exponents of the density profile should depend on the progenitor system and the mechanism of the SN explosion. While the exponent in the range $7\lesssim n \lesssim 10$ has been inferred from simulations for Type Ia SNe~\citep[e.g.,][]{1969ApJ...157..623C,2010ApJ...708.1025K},  the exponent may become smaller for weaker explosions such as Type Iax SNe. Actually, simulations of the mass eruption from a massive star before core-collapse have found a profile of $n\sim6$ \citep[e.g.,][]{2020A&A...635A.127K,2020ApJ...897L..44T}. Given that, we consider a range of $6 \leq n \leq 10$ and set $n = 6$ as the fiducial value. In this case, the radii of the forward shock $r_\mathrm{out,f}$, the contact discontinuity $r_\mathrm{out,c}$, and the reverse shock $r_\mathrm{out,r}$ {}{(see Fig. \ref{fig:ejecta_density})} all evolve with the same power-law in time;  
\begin{equation}
r_{\mathrm{out,}i} = \left({B}_ig/\rho_{\rm ISM}^{1/6}\right)t^{1/2}\ (i={\rm f},\ {\rm c},\ {\rm r}).
\end{equation}
The proportionality factors ${B}_i$ for these radii, as well as the density, pressure, and velocity profiles inside both the shocked ejecta and the shocked ISM, are obtained for a given set of parameters ($E_\mathrm{ej}, M_\mathrm{ej}, n_\mathrm{ISM}$). {Here we use the code developed in \cite{Tsuna19} that calculates the hydrodynamical profiles of the shocked region, for an arbitrary value of the adiabatic index and power-law density profiles of the ejecta and ISM for which a self-similar solution exists.} We assume an adiabatic index of $\gamma = 5/3${}{, because this region consists of approximately adiabatic, non-relativistic ionized gas.}  

\subsubsection{{Angular radius}}
As we have shown in Sec. \ref{sec:X-ray}, the {angular radius} of the outer X-ray nebula is {}{$\theta_{\rm out} = 131^{+2}_{-2 -2}~$arcsec} based on the XMM-Newton observation. In our model, this corresponds to the forward shock radius $r_\mathrm{out,f}$ at $t = 840\,\mathrm{yr}$: 
\begin{equation}
    {}{r_{\rm out,f} = 1.46_{-0.04}^{+0.02}~\pc.}
\end{equation}
Note that the estimated error of $r_{\rm out,f}$ is smaller than the errors of the other parameters, thus we neglect the error hereafter. 
Applying this condition to the self-similar solution described above, we obtain a condition
\begin{equation}
\left(\frac{E_{\rm ej}}{10^{48}\,\erg}\right)^{-1/2} \left(\frac{M_{\rm ej}}{0.5~M_\odot}\right)^{1/6}  \left(\frac{\mu_{\rm ISM}n_{\rm ISM}}{0.1~\percc}\right)^{1/3} = 0.86. \label{eqn:EMn}
\end{equation}

\subsubsection{Emission measure}
The emission measure of the outer X-ray nebula can be described as 
\begin{equation}\label{eq:EM_out}
    \EM_\mathrm{out} = n_{\rm e} n_{\rm ion} V = \frac{4\pi}{m_{\rm u}^2}\left(\frac{1}{\mu_{\rm r}^2}\int_{r_{\rm out,r}}^{r_{\rm out,c}} r^2\rho(r)^2 dr + \frac{1}{\mu_{\rm f}^2}\int_{r_{\rm out,c}}^{r_{\rm out,f}} r^2\rho(r)^2 dr \right).
\end{equation}
The first term in the right-hand side of Eq. (\ref{eq:EM_out}) represents the contribution from the reverse shock or the shocked ejecta; $\mu_\mathrm{r}$ is the mean molecular weight and $\rho(r)$ is the density. The second term represents the contribution from the forward shock, or the shocked ISM, where $\mu_\mathrm{f}$ is the mean molecular weight. Using the shock structure obtained from the self-similar solution, the integrals of Eq. (\ref{eq:EM_out}) can be calculated as
\begin{equation}\label{eq:out_f}
    \int_{r_{\rm out,c}}^{r_{\rm out,f}} r^2\rho^2 dr = 2.15\times10^6\,\g^2\cm^{-3}
    \left(\frac{{}{\mu_{\rm ISM}}n_{\rm ISM}}{0.1\,\percc}\right)^2,
\end{equation}
\begin{equation}\label{eq:out_r}
    \int_{r_{\rm out,r}}^{r_{\rm out,c}} r^2\rho^2 dr = 8.96\times10^6\,\g^2\cm^{-3}
    \left(\frac{{}{\mu_{\rm ISM}}n_{\rm ISM}}{0.1\,\percc}\right)^2.
\end{equation}
Note that they do not depend on either $E_\mathrm{ej}$ or $M_\mathrm{ej}$ for a given $r_{\rm out,f}$. The mean molecular weights, {$\mu_\mathrm{r}$, $\mu_\mathrm{f}$ are evaluated as follows. When the composition is pure metal, which is suggested by our analyses of the observed X-ray spectrum in Sec. \ref{sec:X-ray}, the mean molecular weights can be expressed as follows: in the outer regions}
\begin{equation}
   {} {\mu_{\rm out,pm}^2 = \mu_{\rm out,e,pm}\mu_{\rm out,ion,pm},}
\end{equation}
where 
\begin{equation}\label{eqn:mu_e}
        {}{\frac{1}{\mu_{\rm out,e,pm}} = \sum_i \frac{\langle n\rangle_iX_i}{A_i} = 0.47,}
\end{equation}
\begin{equation}\label{eqn:mu_ion}
    {}{\frac{1}{\mu_{\rm out,ion,pm}} = \sum_i\frac{X_i}{A_i} = 0.060. }
\end{equation}
Here, {the subscript "pm" represents the physical quantities for the case of pure metal abundance,} $X_i$ denotes the mass fraction of element $i$,  $A_i$ its mass number, and $\langle n\rangle_i$ is the average number of free electrons originating from element $i$. We adopt these values shown in table~\ref{tab:mass}. The number of free electrons is obtained from AtomDB 3.0.9 \citep{2012ApJ...756..128F} using the abundance ratios and the ionization degree for each element. We have used $k_\mathrm{B}T_{\rm e}$ and $\tau$ obtained from our X-ray analysis to estimate the degree of ionization. {On the other hand, the mean molecular weight for the solar abundance is as follows:}
\begin{equation}
    {\mu_{\rm out,solar}^2 = \mu_{\rm out,e,solar}\mu_{\rm out,ion,solar},} 
\end{equation}
where 
\begin{equation}
    {\frac{1}{\mu_{\rm out,e,solar}} \approx X\times\frac{1}{1}+Y\times\frac{2}{4} = 0.85,}
\end{equation}
\begin{equation}\label{eq:mu_ion_f}
    {}{\frac{1}{\mu_{\rm out,ion,solar}} \approx X\times\frac{1}{1}+Y\times\frac{1}{4} = 0.78.}
\end{equation}
Here $X$ and $Y$ are the mass fractions of hydrogen and helium, respectively.

\begin{figure}[ht]
    \centering
    \includegraphics[width=0.8\columnwidth]{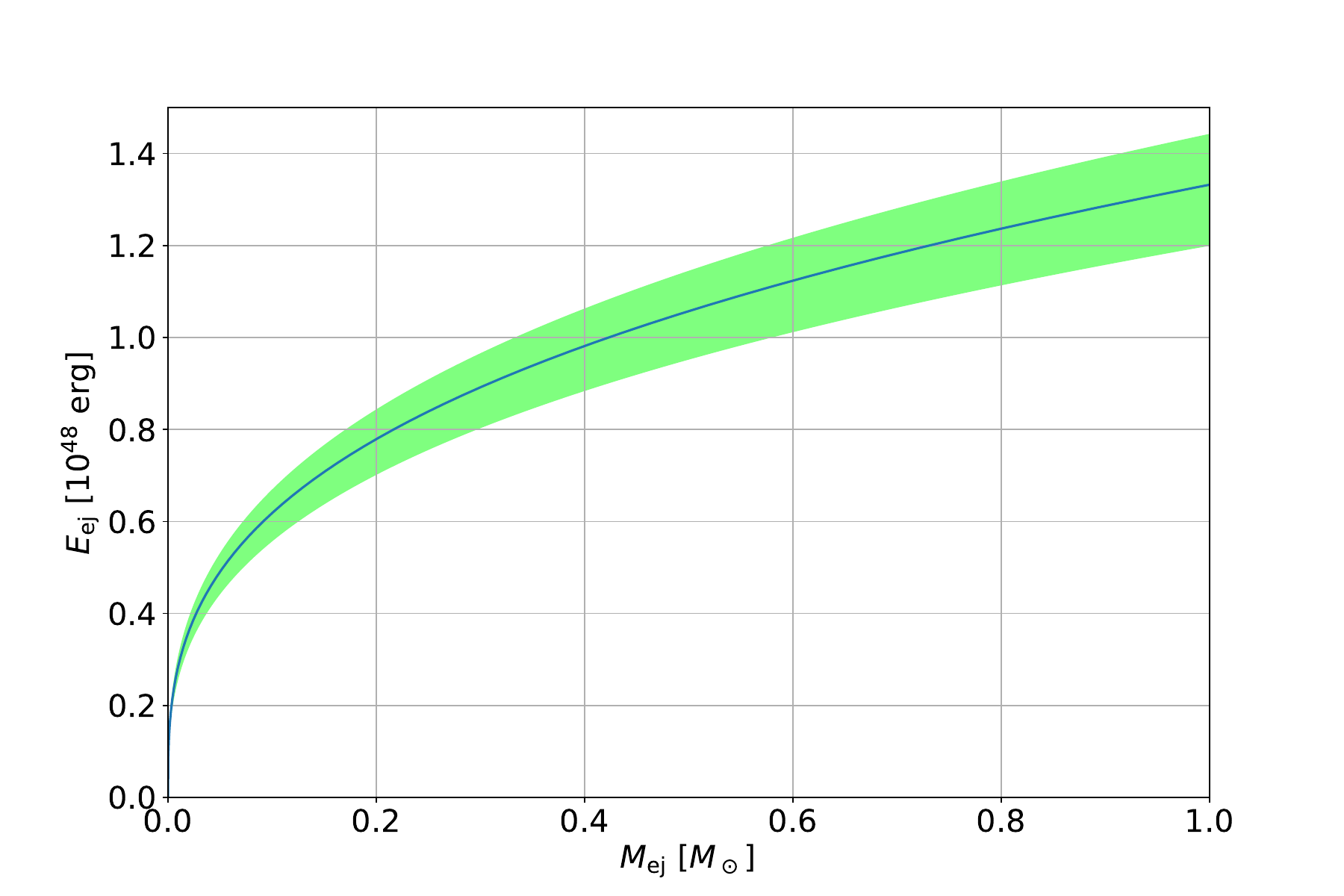}
    \caption{
    The range of the {ejecta kinetic} energy $E_\mathrm{ej}$ and the ejecta mass $M_\mathrm{ej}$ of SN 1181 obtained from our modeling of the outer X-ray nebula of \iras~(Eq. \ref{eqn:EandM}). The blue line and the green shaded region correspond to the best-fit value and 2$\sigma$ uncertainties of the emission measure EM$_\mathrm{out}$ obtained in Sec. \ref{sec:X-ray}. 
    }
    \label{fig:E-M}
\end{figure}

The EM obtained from the X-ray analysis assumes a one-zone uniform shocked gas, while our EM model is a two-zone with different abundances. Though this makes a straightforward comparison difficult, the abundance of the mixture of the two regions is expected to be close to the ISM composition in the above self-similar solution because the mass ratio of the shocked ejecta to the shocked ISM is determined to be $0.28$~{}{\citep[][]{1982ApJ...258..790C}}. Thus, we compare the calculated EM (Eq. \ref{eq:EM_out}, {}{adopting $\mu_{\rm r}=\mu_{\rm out,pm},\ \mu_{\rm f}= \mu_{\rm out,solar}$}) with the value obtained by the X-ray analysis assuming the near-solar abundance ({}{$\EM_{\rm out}= 3.7\pm1.0\times 10^{54}~\percc$, Table \ref{tab:param}}). By substituting Eqs. (\ref{eq:out_f}-\ref{eq:mu_ion_f}) into Eq. (\ref{eq:EM_out}), we can determine the ISM number density as 
\begin{equation}\label{eqn:n_ISM}
   {n_{\rm ISM}} = 0.11~\percc\left(\frac{\EM_{\rm out} }{3.7\times10^{54}~\percc}\right)^{1/2}.
\end{equation}
{}{Here we have used $\mu_{\rm ISM}$ in Eq. (\ref{eqn:mu_ISM})}. Substituting this into Eq. (\ref{eqn:EMn}), we obtain a condition
\begin{equation}\label{eqn:EandM}
    \left(\frac{E_{\rm ej}}{10^{48}\,\erg}\right)^{-1/2} \left(\frac{M_{\rm ej}}{0.5~M_\odot}\right)^{1/6}  = 0.97 \left(\frac{\EM_{\rm out}}{3.7\times 10^{54}~\percc}\right)^{-1/6}.
\end{equation}
Figure \ref{fig:E-M} shows the obtained relation between $E_\mathrm{ej}$ and $M_\mathrm{ej}$. Given that \iras~hosts a WD remnant, the ejecta mass would not be larger than $M_\mathrm{ej} \lesssim 1\,M_\odot$. In this case, the {ejecta kinetic} energy is constrained to be $E_\mathrm{ej} \lesssim 10^{48}\,\mathrm{erg}$ only from the outer X-ray nebula. These results are broadly consistent with the previous work~\citep{Oskinova_et_al_20}.  

\subsection{Inner X-ray nebula}\label{sec:inner}
Next, we consider the inner X-ray nebula. We calculate the position of the contact surface between the shocked wind and shocked ejecta using the thin shell approximation \citep[e.g.,][]{1992ApJ...395..540C}. 
The mass, momentum, and energy conservations of the shell can be described as 
\begin{equation}\label{eqn:shell_3}
    \frac{d}{dt}M_{\rm sh,in} = \dot{M}_{\rm w} + 4\pi r_{\rm sh}^2 \rho_{\rm ej}(r_{\rm sh})\left(v_{\rm sh}+\sqrt{\frac{2GM_*}{r_{\rm sh}}}\right).
\end{equation}
\begin{equation}\label{eqn:shell_1}
    M_\mathrm{sh, in} \frac{dv_\mathrm{sh}}{dt} = 4 \pi r_\mathrm{sh}^2 \left[p_{\rm sh} -\rho_\mathrm{ej}(r_\mathrm{sh}) \left(v_\mathrm{sh}+\sqrt{\frac{2GM_*}{r_{\rm sh}}}\right)^2\right]-\frac{G M_\mathrm{sh, in}M_*}{r_\mathrm{sh}^2},
\end{equation}
\begin{equation}\label{eqn:shell_2}
    \frac{d}{dt}\left[\frac{4\pi r_\mathrm{sh}^3}{3} \frac{p_{\rm sh}}{\gamma-1}\right] = L_\mathrm{w} - p_{\rm sh} \times 4\pi r_\mathrm{sh}^2 v_\mathrm{sh},
\end{equation}
respectively. 
Here $M_\mathrm{sh, in}$, $r_\mathrm{sh}$, $v_\mathrm{sh}$ are the mass, radius, and velocity of the shocked shell, respectively.  The mass of the central star is denoted by $M_*$, $p_{\rm sh}$ is the pressure in the wind bubble, $\rho_\mathrm{ej}$ is the unshocked ejecta density, and $v_\mathrm{ej}$ is the velocity. In Eqs. (\ref{eqn:shell_3}) and (\ref{eqn:shell_2}), we assume that the tail part of the SN ejecta {are} falling back toward the central WD with a velocity of $-\sqrt{2GM_*/r_\mathrm{sh}}$. This can be justified given that the inferred {ejecta kinetic} energy from the outer X-ray nebula, $E_\mathrm{ej} \sim 10^{48}\,\mathrm{erg}$ (see Figure \ref{fig:E-M}), is much smaller than the gravitational binding energy of the central WD, $GM_*^2/R_* \sim 10^{50}\,\mathrm{erg}$~\citep[][]{2019ApJ...887...39K}. {}{Here, $R_*$ refers to the radius of the central WD, whose value is around $10^{9}~\cm$ \citep[][]{2019ApJ...887...39K}. Note that $R_*$ is different from the photospheric radius ($\sim 1\times 10^{10}$ cm) since the wind is optically thick, and most of the WD mass is contained within $R_*$.} The density at the forward shock front is estimated from Eq. (\ref{eqn:rho_ej1}) with fixing $\delta = 1.5$, which is appropriate for a fallback tail~\citep[e.g.,][]{Tsuna2021}\footnote{
Eq. (\ref{eqn:rho_ej1}) assumes that the entire ejecta {are} expanding homologously, which is inconsistent with the assumption that the tail part of the SN ejecta {are} falling back to the central WD. However, both the fallback velocity $\sim \sqrt{2GM_*/r_\mathrm{sh}}$ or the homologous expansion velocity $\sim r_\mathrm{sh}/t$ are basically much smaller than the shock velocity $v_\mathrm{sh}$, and this does not practically affect the results of our calculation.
}.

In order to follow the evolution of the inner nebula, we integrate Eqs. (\ref{eqn:shell_3}), (\ref{eqn:shell_1}), and (\ref{eqn:shell_2}) from $t = t_{\rm w}$ for a given set of ($t_{\rm w}$, $E_{\rm{ej}}$, $M_{\rm{ej}}$). {}{Substituting the value of $\EM_{\rm out}$ measured from the X-ray data for the outer nebula into Eq. (\ref{eqn:EandM})}, we have two independent parameters ($t_{\rm w}$, $M_{\rm{ej}}$). The initial conditions are given as $[v_{\rm sh,0},p_{\rm sh,0}, M_{\rm sh,0}] = [-\sqrt{2GM_*/r_{\rm sh, 0}},3(\gamma-1)L_{\rm w}/4\pi r_{\rm sh, 0}^2 v_\mathrm{w},\int_{R_*}^{r_{\rm sh, 0}} 4\pi r^2 \rho_{\rm ej}(r)dr]$. We set the initial shell radius sufficiently small, $r_{\rm sh, 0} = 10^{10}\,\cm$ so that the initial conditions do not affect the evolution on a timescale of 100 yr. We investigate a range of wind mass loss rate $\dot M_\mathrm{w} = (7.5\mbox{--}40)\times 10^{-7}\,M_\odot\,\mathrm{yr}^{-1}$ \citep[][]{Lykou2022}, inferred from the optical spectroscopic observations.

\subsubsection{{Angular radius}}
\begin{figure}[ht]
    \centering
    \includegraphics[width=0.8\columnwidth]{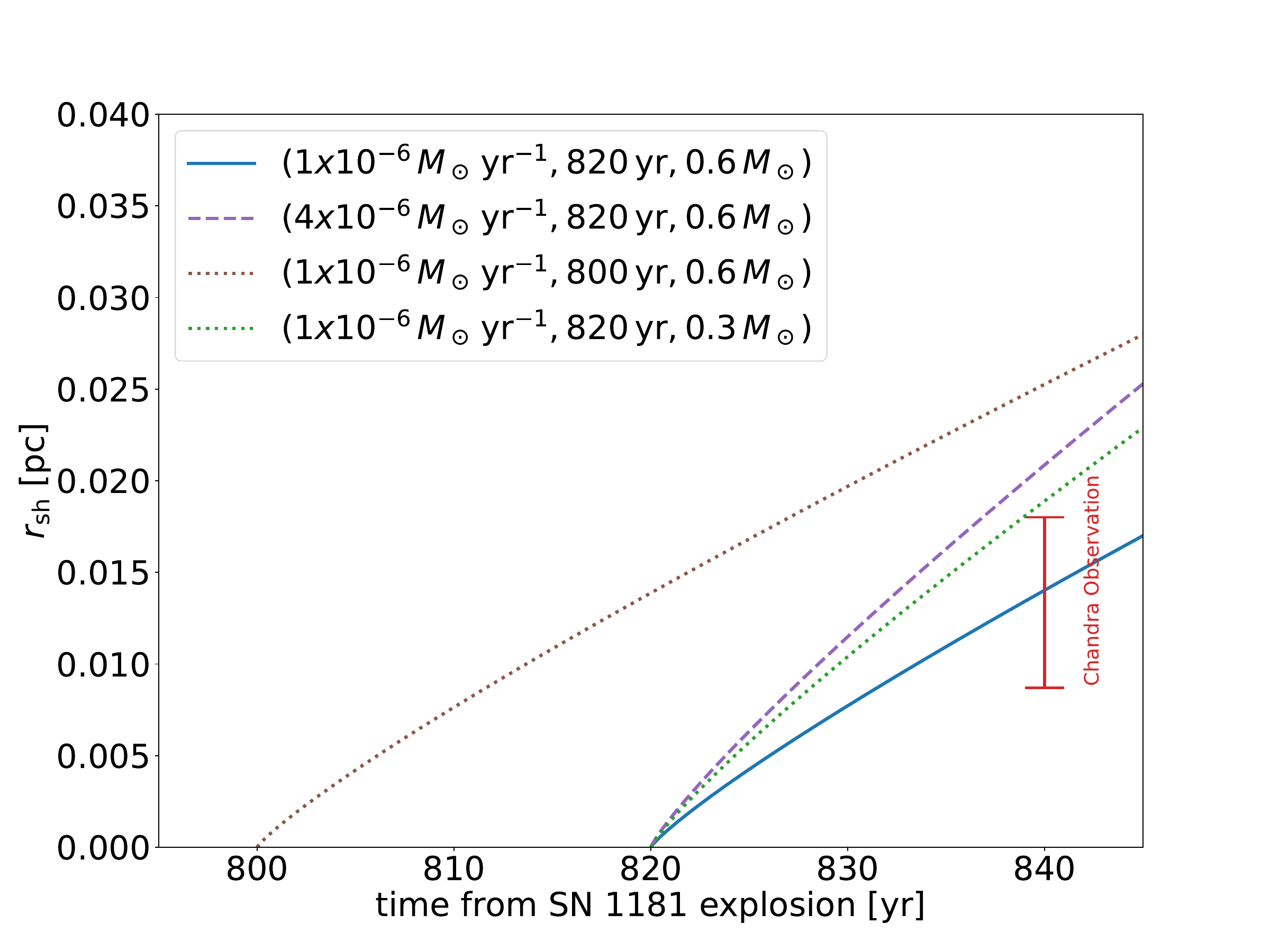}
    \caption{The time evolution of the radius of the inner X-ray nebula $r_\mathrm{sh}$ calculated for several parameter sets $(\dot{M}_{\rm w},t_{\rm w},M_{\rm{ej}})$ using our dynamical model. The red line indicates the value obtained by the Chandra observations. }
    \label{fig:shell_rad}
\end{figure}

As we show in Sec. \ref{sec:X-ray}, the {angular radius} of the inner X-ray nebula is constrained as $0.77\,\mathrm{arcsec} < \theta_{\rm in} < 1.6$ arcsec based on the Chandra observation in 2021. In our model, this corresponds to the thin shell radius, i.e. the radius of the wind termination shock $r_\mathrm{sh}$ at $t = 840\,\mathrm{yr}$:  
\begin{equation}\label{eqn:r_constaraint}
0.0087~\pc<r_{\rm sh}<0.018~\pc.
\end{equation}
Figure \ref{fig:shell_rad} shows the time evolution of the inner nebula radius calculated for several parameter sets. 
One can see that $r_{\rm sh}$ at $t = 840\,\mathrm{yr}$ becomes larger for a smaller $M_{\rm ej}$, a larger $\dot{M}_{\rm w}$, and/or smaller $t_\mathrm{w}$. This is simply because the wind keeps pushing the fallback ejecta since its {launch}. Therefore, too small $M_{\rm ej}$  or too small $t_{\rm w}$  leads to a too-expanded shell, which is inconsistent with the Chandra observation. This can give a constraint on the parameter set $(\dot{M}_{\rm w},t_{\rm w},M_{\rm{ej}})$.

\subsubsection{Emission measure}\label{sec:em}
\begin{figure}[ht]
    \centering
    \includegraphics[width=0.8\columnwidth]{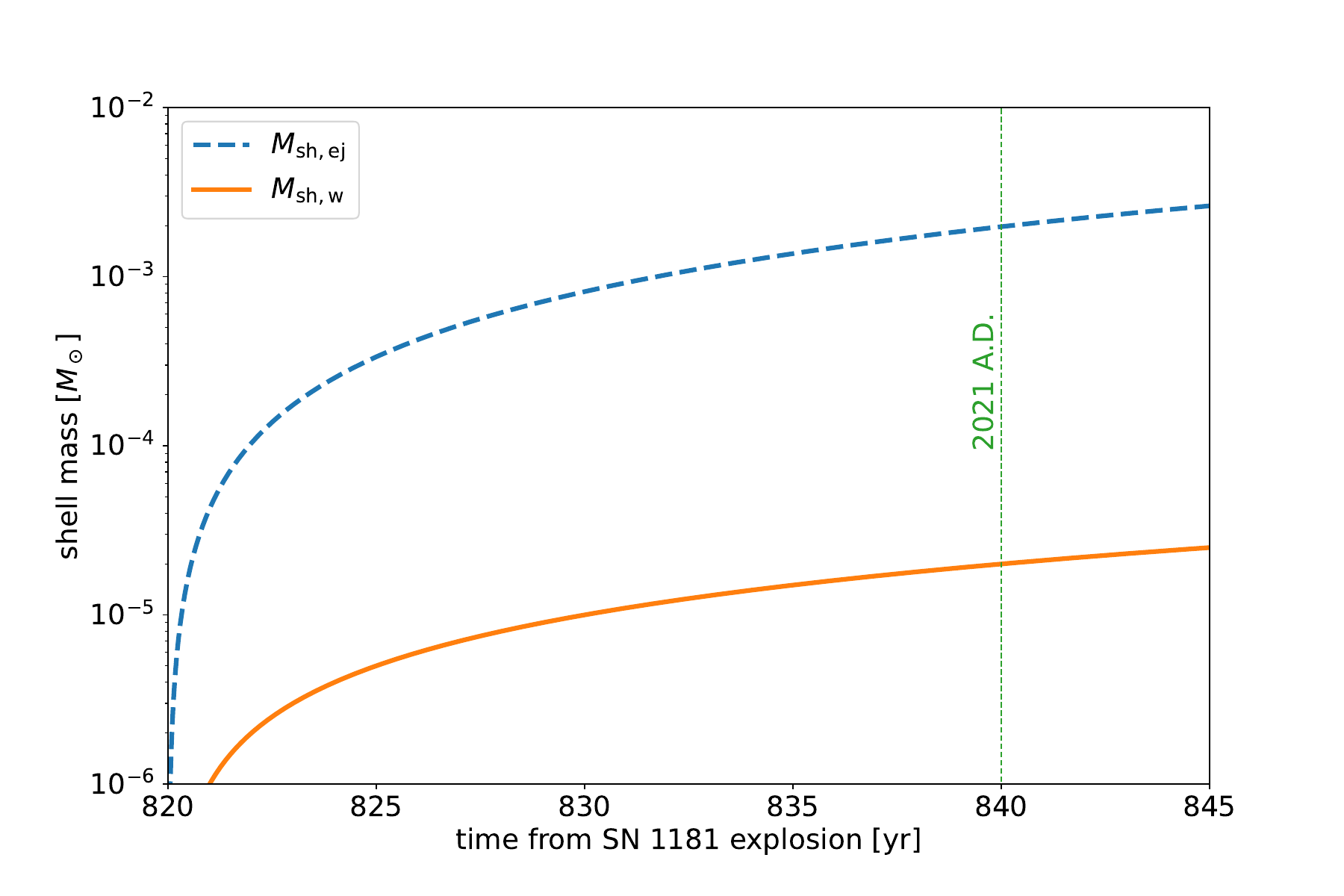}
    \caption{The time evolution of the mass in the shocked wind (red solid line) and the shocked SN ejecta (blue dashed line) of the inner X-ray nebula, calculated for a parameter set $(\dot{M}_{\rm w},t_{\rm w},M_{\rm{ej}}) = (1\times10^{-6}~M_\odot\,\yr^{-1},820~\yr,0.3~M_\odot)$ using our dynamical model. The green dotted line indicates the year 2021 AD when the X-ray observations were done. }
    \label{fig:shell_evo}
\end{figure}

By modeling the expansion of the inner nebula, we can also estimate the total mass in the shocked region. Figure \ref{fig:shell_evo} shows the time evolution of the shocked mass for a case with $(\dot{M}_{\rm w}, t_{\rm w},M_{\rm{ej}}) = (1\times10^{-6}~M_\odot\,\yr^{-1}, 820~\yr, {}{0.3}~M_\odot)$. The solid and dashed lines indicate the contribution from the shocked wind swept by the reverse shock and the shocked SN ejecta swept by the forward shock, respectively. One can see that the latter is 2-3 orders of magnitude larger than the former. Importantly, however, the shocked SN ejecta component will not contribute to the emission measure of the inner X-ray nebula; {}{as we show in Appendix \ref{app:ejecta}, the electron temperature is kept below $3\mbox{--}4\,\mathrm{eV}$ in the shocked SN ejecta.} Thus, we only consider the shocked wind as the source of the inner X-ray nebula. In this case, the emission measure of the inner nebula can be estimated as 
\begin{equation}\label{eqn:EM_in_w}
    \EM_{\rm in} = n_{\rm e} n_{\rm ion} V_{\rm sh,w}  = \frac{M_{\rm sh, w}^2}{\mu_{\rm e}\mu_{\rm ion}m_{\rm u}^2V_{\rm sh,w}}, 
\end{equation}
where $M_\mathrm{sh, w} = \dot M_\mathrm{w} (t - t_\mathrm{w})$ is the total mass of the shocked wind, and 
\begin{equation}\label{eqn:wind_volume}
    V_\mathrm{sh,w} \approx \frac{4}{3}\pi\left(r_\mathrm{sh}^3-r_\mathrm{in,r}^3\right)=\frac{4}{3}\pi r_\mathrm{sh}^3\left[ 1-\left(\frac{2}{1+\gamma}\right)^3\right]
\end{equation}
is the volume of the shocked wind region. Here we estimate the positions of the reverse shock fronts as 
\begin{equation}\label{eq:r_in_r}
    r_\mathrm{in, r} \approx (1-{\cal C}) r_\mathrm{sh} = \frac{2}{1+\gamma} r_\mathrm{sh}.
\end{equation}
With an approximation that the shock region is plane-parallel and adiabatic, the compression ratio is ${\cal C} = (\gamma-1)/(\gamma+1)=4$ for $\gamma = 5/3$ in the strong shock limit. {Since we consider the inner X-ray region to be formed by the interaction between the C-O-rich wind from the central WD and the unshocked ejecta, we consider the abundance in this region to be pure metal, and adopt the value of $\EM_{\rm in}$ obtained from the analysis under the pure-metal assumption.} The mean molecular weights in the shocked wind region are calculated for pure metal abundance with ionization inferred from the X-ray spectroscopy in Sec. \ref{sec:X-ray}: 
\begin{equation}
        {}{\frac{1}{\mu_{\rm in,e,pm}} = \sum_i \frac{\langle n \rangle_i X_i }{A_i} = 0.49,}
\end{equation}
\begin{equation}
    {}{\frac{1}{\mu_{\rm in,ion,pm}} = \sum_i \frac{X_i }{A_i}\sim 0.062.}
\end{equation}

\begin{figure}[ht]
    \centering
    \includegraphics[width=0.8\columnwidth]{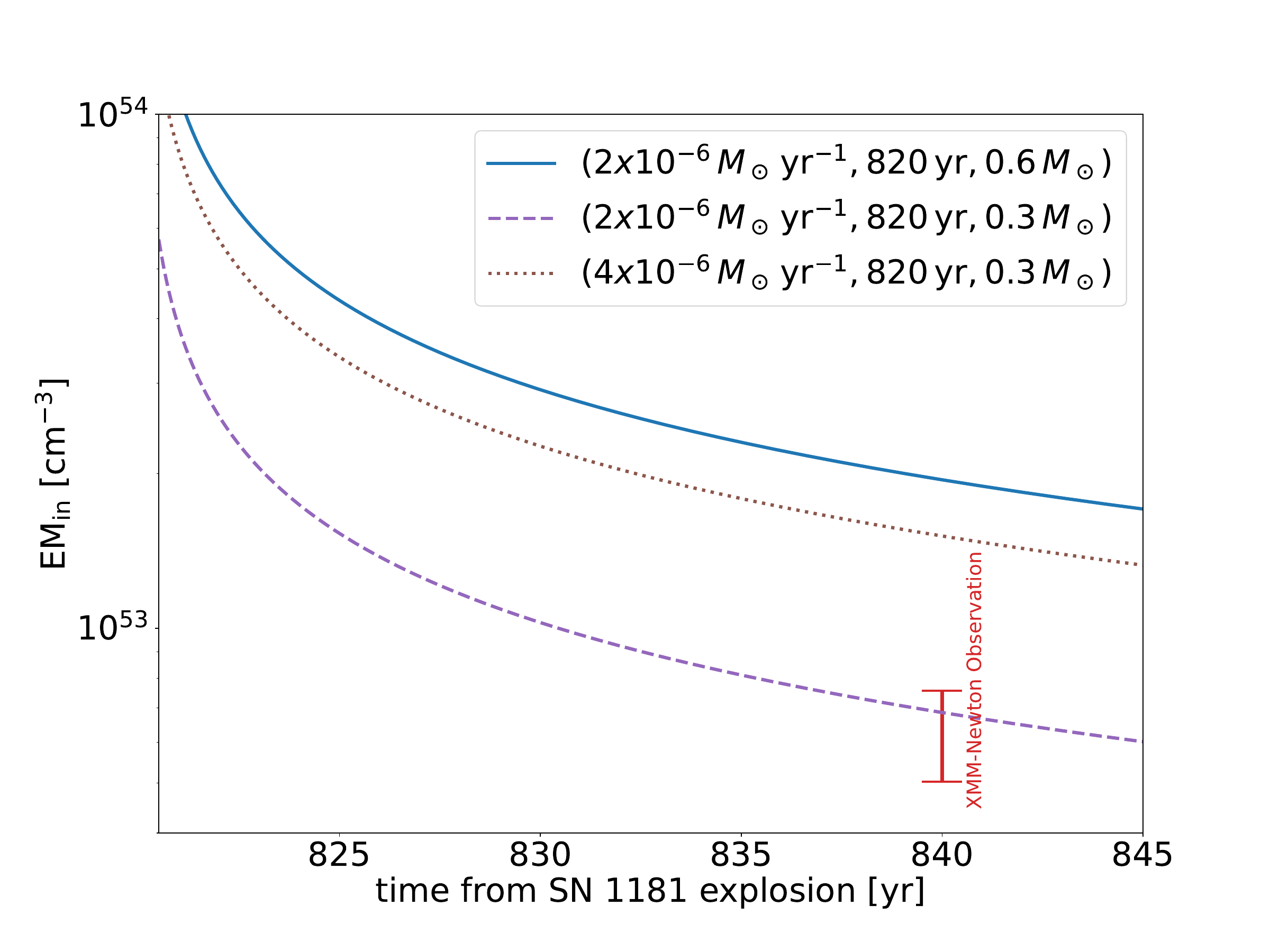}
    \caption{The time evolution of the emission measure of the inner X-ray nebula EM$_\mathrm{in}$ calculated for several parameter sets $(\dot{M}_{\rm w},t_{\rm w},M_{\rm{ej}})$ using our dynamical model. The red {bar} indicates the value obtained by {our spectral analysis of the XMM-Newton observations in Sec. \ref{sec]spectral_prop} .}}
    \label{fig:EM_evo}
\end{figure}

By comparing the calculated EM$_\mathrm{in}$ of the shocked wind (Eq. \ref{eqn:EM_in_w}) with the observed value for the pure metal model (Tables \ref{tab:specb} and \ref{tab:param}), we can obtain an additional constraint on our model parameters. {}{Here, as shown in Table~\ref{tab:param} the value of $\EM_{\rm in}$ is constrained from the X-ray analysis as follows:}
\begin{equation}\label{eqn:EM_constaraint}
5.0\times10^{52}~\percc < \EM_{\rm in} < 7.6 \times10^{52}~\percc
\end{equation}
Figure \ref{fig:EM_evo} shows the time evolution of EM$_\mathrm{in}$ calculated for several parameter sets {}{$(\dot{M}_{\rm w}, t_{\rm w},M_{\rm{ej}}) = (2\times 10^{-6}\,M_\odot\,\mathrm{yr^{-1}},820\,\mathrm{yr}, 0.6\,M_\odot),(2\times 10^{-6}\,M_\odot\,\mathrm{yr^{-1}},820\,\mathrm{yr}, 0.3\,M_\odot),(4\times 10^{-6}\,M_\odot\,\mathrm{yr^{-1}},820\,\mathrm{yr}, 0.6\,M_\odot)$}. As expected, a larger $\dot M_\mathrm{w}$ gives a larger shocked mass and thus a larger EM$_\mathrm{in}$. In addition, a larger $M_\mathrm{ej}$ also gives a larger EM$_\mathrm{in}$; this is because the inner nebula expansion becomes slower for a larger $M_\mathrm{ej}$ and thus the volume of the inner nebula $V_\mathrm{sh, w}$ becomes smaller (see Eq. \ref{eqn:EM_in_w}).

\subsection{Model parameter determination}\label{sec:result}
Combining all the conditions shown in the previous subsections, we here determine the parameters of our dynamical model. 
First, we constrain the ISM density $n_{\rm ISM}$ (Eq. \ref{eqn:n_ISM}) and obtain a relation between $E_{\rm {ej}}$ and $M_{\rm {ej}}$ (Eq. \ref{eqn:EandM}) from the observed size and EM of the outer X-ray nebula. Then, we search sets of the residual parameters ($t_\mathrm{w}$, $M_\mathrm{ej}$) that can consistently explain the observed size and EM of the inner X-ray nebula. In doing so, we allow the wind mass loss rate $\dot M_\mathrm{w}$ to vary in the range inferred from the optical spectroscopy. {}{In addition, we perform parameter search ($t_\mathrm{w}$, $M_\mathrm{ej}$) in the range of $12~\yr<\Delta t_{\rm w} < 840~\yr$ and $M_{\rm ej} < 3~M_\odot$. Here, $\Delta t_\mathrm{w}=840~\yr-t_{\rm w}~[\yr]$ refers to the timing of the wind launch. The lower limit of $\Delta t_{\rm w}>12~\yr$ comes from the constraints of Swift XRT detection in 2009 A.D., which identified an X-ray counterpart comparably bright as the currently observed inner X-ray nebula~\citep[][]{2013yCat.9043....0E,Lykou2022}. We conservatively adopt a loose upper limit of the ejecta mass $M_{\rm ej} < 3~M_\odot$, as it cannot be heavier than twice the Chandrasekhar limit for a WD merger.}

{}{In this work, we search for the possible range of parameters that satisfy the observational constraint (Eqs. \ref{eqn:r_constaraint} and \ref{eqn:EM_constaraint}). Note that we do not conduct any optimization methods, and therefore we aim to obtain the allowed range of parameters rather than a statistical constraint on them.}
\begin{figure}[ht]
    \centering
    \includegraphics[width=0.8\columnwidth]{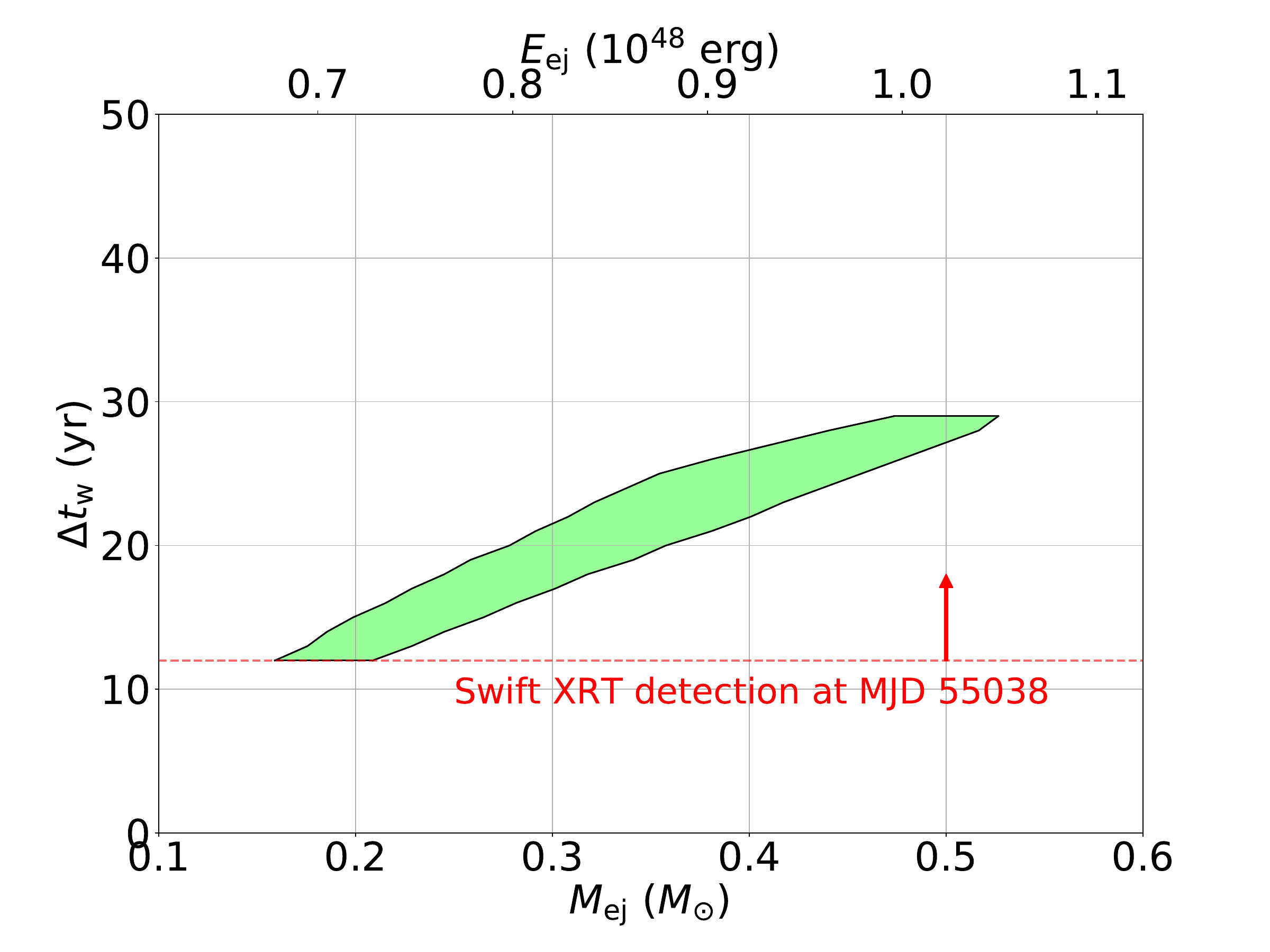}
    \caption{Constraints on the parameters of our dynamical model. The shaded region indicates the allowed range of the parameters with respect to the ejecta mass (lower horizontal axis) and the {ejecta kinetic} energy (upper horizontal axis) of SN 1181, and the timing of the launch of the currently observed wind (vertical axis). The relation between the upper and lower horizontal axis is based on Eq. (\ref{eqn:EandM}). The red dashed line indicates the timing of the XRT observation of \iras.}
    \label{fig:result_all}
\end{figure}

Figure \ref{fig:result_all} shows the allowed range of the parameters with respect to the SN ejecta mass $M_\mathrm{ej}$ (lower horizontal axis), the SN {ejecta kinetic} energy $E_\mathrm{ej}$ (upper horizontal axis), and the timing of the wind launch $\Delta t_\mathrm{w}=840~\yr-t_{\rm w}~[\yr]$ (vertical axis) consistent with the X-ray observations by XMM-Newton and Chandra. 
We find that the size constraint of the inner nebula is tight; to confine the wind termination shock as observed, it only allows a very recent launch of the wind and a sufficiently large ejecta mass.  On the other hand, the combination of the emission measures of the inner and outer nebulae set an upper limit on the ejecta mass. {The earlier Swift XRT detection is considered as a lower limit for $\Delta t_\mathrm{w}$ in our models.} In summary, we obtain the following constraints on the model parameters:\footnote{{ROSAT detected X-rays with a comparable flux from the same source in 1990 A.D. \citep{2000IAUC.7432....3V}, but it is uncertain whether the source was extended or not. In other words, the X-rays detected by ROSAT  might not originate from the central point-like source. Thus we do not use this detection as a constraint on $\Delta t_{\rm w}$ or $t_{\rm w}$.}}
\begin{equation}\label{eq:fit_nism}
    0.097~\percc < n_\mathrm{ISM} < 0.13~\percc,
\end{equation}
\begin{equation}
    0.77\times10^{48}~\erg< E_{\rm ej} < 1.1\times10^{48}~\erg,
\end{equation}
\begin{equation}\label{eqn:result_Mej}
    0.18~M_\odot < M_{\rm ej} <  0.53~M_\odot,
\end{equation}
and 
\begin{equation}\label{eqn:dt_constaraint}
    12~\yr < \Delta t_{\rm w} <30~\yr,
\end{equation}
or
\begin{equation}\label{eqn:alpha_constaraint}
   810~\yr < t_{\rm w} < 828~\yr.
\end{equation}

\subsection{Consistency check with other observations}\label{sec:consist}
The model parameters (Eqs. \ref{eq:fit_nism}-\ref{eqn:alpha_constaraint}) are obtained by mainly fitting the {angular radius} and the emission measures of the X-ray nebulae. We here test the validity of the fitted model by comparing it with other relevant observations of \iras.

\subsubsection{The plasma properties in the X-ray nebulae}\label{sec:uncertain}

In addition to measuring {angular radius} and emission measures, our X-ray {spectral} analysis has determined relevant quantities of the X-ray-emitting plasma, such as ionization timescale and electron temperature (see Table \ref{tab:specb}). The ionization timescale is estimated as $\tau \approx n_\mathrm{e} \times \Delta t$, where $n_\mathrm{e}$ is the {post-shock} electron number density and $\Delta t$ is the dynamical time of the plasma, and it provides an indicator of the degree of ionization.

{In the allowed parameter space obtained by our analysis, the ionization timescale in the inner shocked region can be estimated as }$\tau_\mathrm{in} \approx\rho_\mathrm{sh,w}/(\mu_\mathrm{e}m_\mathrm{u}) \times \Delta t_\mathrm{w} \sim 10^{10\mbox{--}11}\,\mathrm{s\,cm^{-3}}$. {Here, $\rho_\mathrm{sh,w}$ is the density of the inner shocked region obtained by $\rho_\mathrm{sh,w} = M_\mathrm{sh,w}/V_\mathrm{sh,w}.$}
{The ionization timescale in the outer shocked region has a value of} $\tau_\mathrm{out} \approx 4 n_\mathrm{ISM} \times 840\,\mathrm{yr} \sim 10^{10}\,\mathrm{s\,cm^{-3}}$. Both timescales are broadly consistent with the values obtained from the X-ray spectroscopy. In order to achieve CIE in the nebulae with the estimated temperatures, $\tau \gtrsim 10^{11}\,\mathrm{s\,cm^{-3}}$ is required for oxygen and $\gtrsim 10^{12}\,\mathrm{s\,cm^{-3}}$ for heavier elements~\citep[e.g.,][]{2010ApJ...718..583S}. Hence we infer that both the inner and outer nebulae are likely not in CIE. To estimate the electron temperature accurately, a detailed plasma calculation coupled with shock dynamics is required~\citep[e.g.,][]{Hamilton83,Masai94,2021ApJ...914...64T}, which is beyond the scope of this paper. Still, if the shock downstream is adiabatic and the electrons and ions are far from CIE, which are good approximations, particularly for the shocked wind or the inner nebula, the electron temperature can be estimated as $k_{\rm B}T_\mathrm{e, in} \approx 511\,\mathrm{keV}\times (v_\mathrm{w}/c)^2 \sim 1.3\,\mathrm{keV}$, which is also consistent with X-ray spectroscopy. Here $c$ denotes the speed of light.

\subsubsection{The infrared observations}
Infrared observations of \iras~revealed a dust-rich ring with an {angular radius} of $\theta_\mathrm{d} \sim 1$ arcmin or a physical size of $r_\mathrm{d} \sim 0.66\,\mathrm{pc}$. {The ring is surrounded by a diffuse infrared halo~\citep{2019Natur.569..684G,Lykou2022}, whose expansion velocity was assumed to be similar to that of the shocked ejecta $v_\mathrm{SII} \sim 1,100\,\mathrm{km\,s^{-1}}$ based on the [SII] line signatures observed for $\theta_\mathrm{SII} \sim 90\,\mathrm{arcsec}$ or physical size of $r_\mathrm{SII} \sim 0.99\,\mathrm{pc}$~\citep{2021ApJ...918L..33R,Lykou2022,2023ApJ...945L...4F}. }

In our model, we expect that there are relatively low-temperature layers in the unshocked region, $r_\mathrm{sh} < r < r_\mathrm{out, r}$, where the reverse shock radius is calculated as $r_\mathrm{out, r} = 1.04\,\mathrm{pc}$. Note that the ratio between the forward shock radius and the reverse shock radius is $r_\mathrm{out, f}/r_\mathrm{out, r} = 1.39$, according to the self-similar solution with $n = 6$ and $s = 0$~\citep{1982ApJ...258..790C}. Therefore, we have $r_\mathrm{sh} < r_\mathrm{d}, r_\mathrm{SII} < r_\mathrm{out, r}$, which is consistent with the observed sizes of the dust-rich ring and the diffuse infrared halo. 
Since the inferred expansion velocity of the outer edge of the unshocked region is {}{$\sim r_\mathrm{out,r}/840\,\mathrm{yr} \sim 1,200\,\mathrm{km\,s^{-1}}$}, which is comparable to $v_\mathrm{SII} = 1,100\,\mathrm{km\,s^{-1}}$, the region with the [SII] line signature should be close to the reverse shock front. On the other hand, the outer edge of the dust-rich ring, which is at a slightly smaller radius, may correspond to the sublimation front of the dust that is irradiated by the radiation from the reverse shock. Although a dynamical calculation including the formation and destruction of dust is beyond the scope of this paper, it is important, particularly in light of the recently identified radially aligned filamentary structure around this region \citep{2023ApJ...945L...4F}.

\section{Summary and Discussion}\label{sec:discussion}\label{sec:conclusion}
In this paper, we construct a dynamical model for \iras~to determine the physical parameters of the system and consistently explain the multi-wavelength data. We first analyze the archival X-ray data obtained by XMM-Newton and Chandra to extract information about the central (point-like) component and diffuse components, such as their {angular radius}, emission measures, ionization timescales, and electron temperatures. In particular, we confirm from the Chandra data that the central component has a finite {angular radius} of $\theta_{\rm in} = 0.77\mbox{--}1.6~$arcsec or a physical size of {}{$r_\mathrm{sh} = 0.0087\mbox{--}0.018\,\pc$}. Assuming that \iras~is a remnant of SN 1181, we interpret that the central component originates from the shocked region between the carbon-burning-ash enriched wind from the central WD J005311 and the SN ejecta, while the diffuse component corresponds to the shocked region where the SN ejecta collide with the ISM. Based on this picture, we construct a dynamical evolution model for the inner and outer shocked regions and find that the X-ray properties of \iras~can be reproduced by a case with an SN {ejecta kinetic} energy of $E_\mathrm{ej} = (0.77\mbox{--}1.1)\times 10^{48}\,\mathrm{erg}$, an SN ejecta mass of $M_\mathrm{ej} = 0.18\mbox{--}0.53\,M_\odot$, if the currently observed intense wind started to blow a few decades ago, $\Delta t_\mathrm{w} = 12\mbox{--}30\,\mathrm{yr}$. In other words, the wind began blowing approximately after A.D. 1990. We have confirmed that our model is also consistent with the infrared geometry, including the dust-rich ring. In the following sections, we discuss the implications of these results for the progenitor system and the evolution of the remnant including the mechanism of the wind.

\subsection{The progenitor system}
The SN parameters obtained by our dynamical model are broadly consistent with previous independent estimates: 
{\cite{Oskinova_et_al_20} estimated the mass of the X-ray emitting gas as $\sim 0.1\,M_\odot$ from the XMM-Newton data, while \cite{Lykou2022} assessed the mass of ionized gas based on the tentative H$\alpha$ detection, concluding it to be $0.15\pm0.05~M_\odot$, both of which are considered lower limits of the SN ejecta mass. {}{In our model, this tentative H$\alpha$ detection reported by \cite{Lykou2022} corresponds to our outer shocked ISM region.}
By assuming an expansion velocity of $v_\mathrm{SII} \sim 1,100\,\mathrm{km\,s^{-1}}$, the ejecta kinetic energy was estimated to be a few $\times 10^{48}\,\mathrm{erg}$~\citep{Lykou2022}.  
These SN parameter values, coupled with the ejecta's metal abundance inferred from the observed X-ray spectra, suggest that the explosion corresponds to a weak type of thermonuclear event from a degenerate system, consistent with type Iax SNe~\citep[e.g.,][]{2013ApJ...767...57F,2017hsn..book..375J}}. 
{}{This suggestion is consistent with the comment by \cite{2023ApJ...945L...4F} that the nebula is metal-rich and H-poor.}
Our study, {which newly incorporates the effects of the strong wind on the ejecta and nebula dynamics}, further supports the assertion that \iras~is the remnant of Type Iax SN 1181, which emerged approximately 840 years ago.

Type Iax SNe are considered to be produced either from single \citep[e.g.,][]{2005ApJ...632..443L,2013MNRAS.429.2287K,2014MNRAS.438.1762F} or double \citep[e.g.,][]{2018ApJ...869..140K} degenerate systems, {although other possibilities, such as core-collapse systems \citep[e.g.,][]{2010ApJ...719.1445M} and core degenerate systems \citep[e.g.,][]{2018MNRAS.480.4519C,2019NewAR..8701535S}, have also been proposed.}
Since a stellar companion of \WD~has not been identified, the double degenerate merger model is likely for SN 1181. From the observed wind properties, the mass of the remnant \WD~has been estimated to be $M_* \gtrsim 1.1\,M_\odot$~\citep{2019Natur.569..684G,2019ApJ...887...39K}. In this case, the total mass of the progenitor binary, $\sim M_* + M_\mathrm{ej}$, is comparable to or exceeds the Chandrasekhar limit. To produce a Type Iax SN from such a system, a binary consisting of a primary WD with $M_1 \gtrsim 1.1,M_\odot$ would be preferred~\citep{2015ApJ...805L...6S}. 
{This primary WD resides in the heavier end of the WD mass distribution~\citep{2007MNRAS.375.1315K,2018MNRAS.479L.113K} and is likely to originate from an intermediate-mass star with $M_* \gtrsim 5\,M_\odot$~\citep[e.g.,][]{Cummings2018,2020A&A...636A..31T}.}
Systematic studies of double degenerate merger simulations aiming to reproduce \iras~are desirable, for which our model predictions, i.e., $E_\mathrm{ej}$, $M_\mathrm{ej}$, and the ejecta density profile, will be useful.

\subsection{The remnant evolution}
If \iras~is the remnant of SN 1181, the age of \WD~is approximately 840 years. Stellar evolution calculations have shown that the observed properties of \WD~and the surrounding infrared halo can be consistent with a remnant of either CO + CO or ONe + CO binary merger occurring $\sim 1,000\mbox{--}10,000$ years ago~\citep{2016MNRAS.463.3461S,2023MNRAS.524.1031Y,2023ApJ...944L..54W}. Note that, even in the case of CO + CO binaries, the remnant can become an ONe WD as a result of an off-center carbon burning.

An unsettled point is the origin and mechanism of the currently observed wind. We have shown that, in order to be consistent with the size constraint on the inner X-ray nebula obtained by Chandra, the wind started blowing only a few decades ago. Before that, the mass loss rate and/or the wind velocity should have been significantly smaller than those today. Given the remnant age of $\mathcal{O}(10^3)$ yr, the timing of the wind launch appears to be finely tuned. Carbon burning is currently ongoing around the surface of \WD~since the wind is enriched with the ashes. Considering the progenitor system shown in the previous section, the fuel of the wind can be unburned carbon accreted from the secondary CO WD at the merger, and the launch of the wind corresponds to the onset of ``shell'' burning of the fuel triggered by the Kelvin-Helmholtz contraction of the ONe core of \WD, as predicted by the evolution calculations of double degenerate merger remnants~\citep{2016MNRAS.463.3461S}. More dedicated studies on the timing of the wind launch will provide us with hints for understanding the mechanism of the wind and the fate of \WD, whether it will collapse into a neutron star or not.
\subsection{Possible Caveats}
{}{We conclude by briefly discussing several possible caveats of our work.}
{}{We have calculated the evolution of both the outer and inner shocks using a $1$D model under the assumption of spherical symmetry. Though observations indicate that the overall shape of both regions is indeed spherical, filamentary structures are reported by \citet{2023ApJ...945L...4F}. Therefore, it is necessary to include multidimensional effects in the calculation in future works. In addition, to reveal the origin of asphericity, radio observations with good spatial resolutions are required to see the shape of the termination shock of the wind in the future \citep{2024PASJ..tmp...27K}.}

{}{The mechanism of the fast wind is still unknown. We also suggest that there may be a relation between the fact that the wind started blowing recently and the fact that the optical brightness has been decreasing over the last 100 years or so \citep{2023MNRAS.523.3885S}. The wind mechanism and its relation to the luminosity decrease should be investigated in the future.}
\begin{acknowledgments}
\red{This research has made use of data and software provided by the High Energy Astrophysics Science Archive Research Center (HEASARC), which is a service of the Astrophysics Science Division at NASA/GSFC}. This work was financially supported by Japan Society for the Promotion of Science Grants-in-Aid for Scientific Research (KAKENHI) Grant Numbers JP21J00031 (HS), JP20K04010 (KK), JP20H01904 (KK), JP22H00130 (KK), JP22H01265 (HU), JP19H01936 (TT), JP21H04493 (TT), JP20K14512 (KF), JP23H01211 (AB), JP22K03688 (TS), JP22K03671 (TS), and JP20H05639 (TS). DT is supported by the Sherman Fairchild Postdoctoral Fellowship at the California Institute of Technology. TK is supported by RIKEN Junior Research Associate Program.
\end{acknowledgments}

\vspace{5mm}
\facilities{Chandra (ACIS), XMM-Newton (MOS, pn, RGS)
}

\software{HEASoft (v6.20; \citealt{heasarc14}); CIAO (v4.15; \citealt{fruscione06}); SAS (v19.1.0; \citealt{gabriel04})
}

\appendix
\section{Emission from the inner shocked ejecta}\label{app:ejecta}
In this appendix, we show that the electron temperature in the shocked ejecta is kept below $\lesssim 3\mbox{--}4$ eV, and thus, although the mass is significantly larger than the shocked wind, the shocked ejecta {do} not contribute to the emission measure of the inner X-ray nebula. 

\begin{figure}[ht]
    \centering
    \includegraphics[width=0.8\columnwidth]{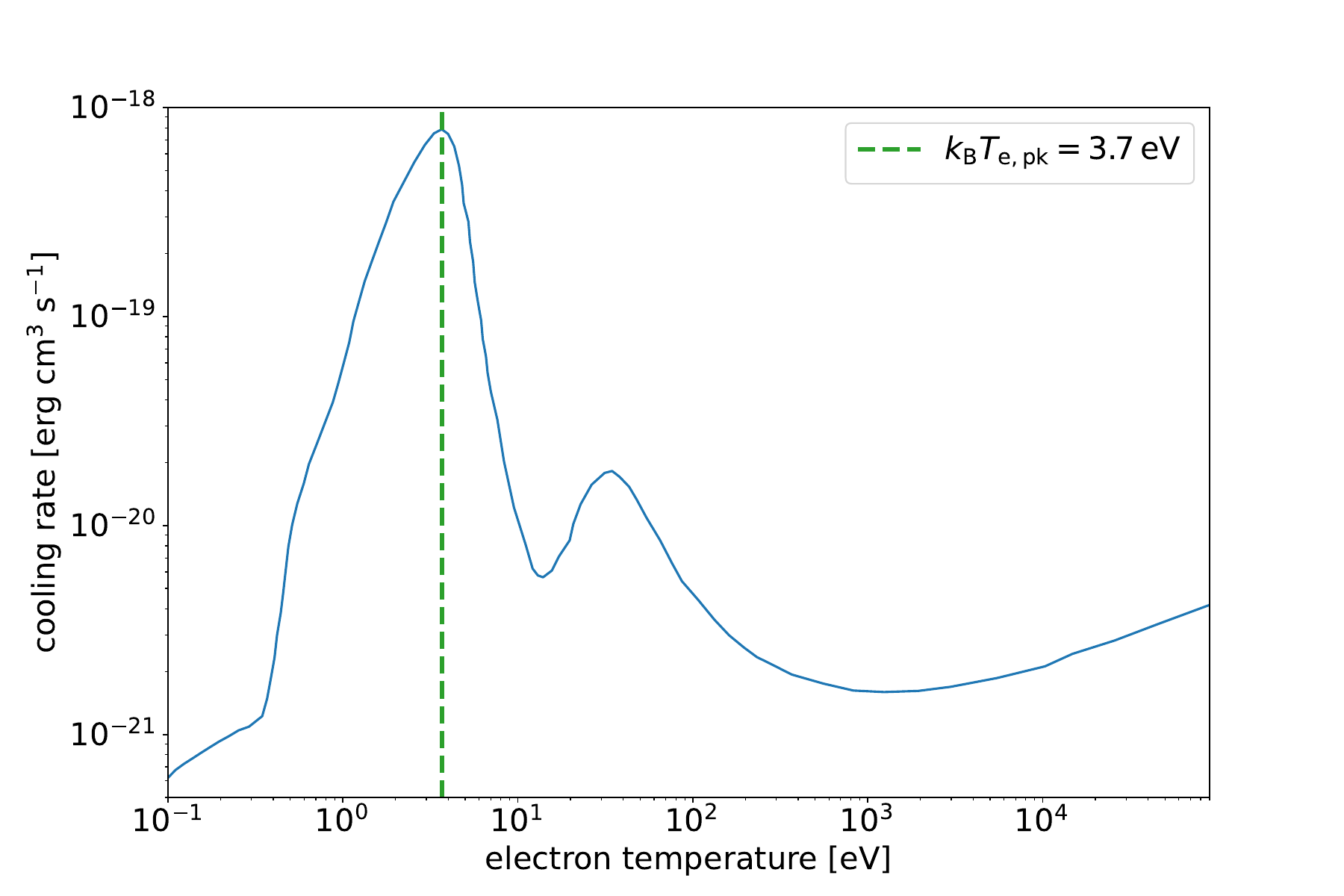}
    \caption{Cooling function of an oxygen dominated gas from \cite{2019MNRAS.489.4465K}.}
    \label{fig:cooling_func}
\end{figure}

At the immediate shock downstream, electrons and ions are first independently heated by collision to the following temperatures respectively.
\begin{equation}
    {}{k_\mathrm{B} T_\mathrm{e,down}(t) = \frac{3}{16}m_ \mathrm{e}\left(v_\mathrm{sh}(t)-\frac{r_\mathrm{sh}(t)}{t}\right)^2,}
\end{equation}
\begin{equation}
    {}{k_\mathrm{B} T_\mathrm{ion,down}(t) = \frac{3}{16}m_\mathrm{ion}\left(v_\mathrm{sh}(t)-\frac{r_\mathrm{sh}(t)}{t}\right)^2.}
\end{equation}
For $t \sim 840\,\mathrm{yr}$, these temperatures are determined from the observed size of the inner nebula, and $k_\mathrm{B} T_\mathrm{e,down} \sim 1\,\mathrm{eV}$ and $k_\mathrm{B} T_\mathrm{ion,down} \sim 20~\mathrm{keV}$, {}{where we adopt $v_\mathrm{sh}=1100~$km/s and $r_\mathrm{sh}=0.018~$pc for simplicity.} Then, the electrons experience collisional heating by the ions and radiative cooling through recombinations of ions and bremsstrahlung radiation. The corresponding heating and cooling timescales can be estimated as 
\begin{equation}\label{eq:t_eq}
    {}{t_{\rm eq}(t) \approx 0.147 \times \frac{A_{\rm ion}A_{\rm e}}{n_{\rm e}Z_{\rm ion}^2Z_{\rm e}^2}\left(\frac{T_{\rm ion,down}(t)}{A_{\rm ion}}+\frac{T_{\rm e,down}(t)}{A_{\rm e}}\right)^{3/2}~\sec},
\end{equation}
\begin{equation}\label{eq:t_cool}
    {}{t_{\rm cool}(t) \approx \frac{k_B T_{\rm e,down}(t)}{n_{\rm ion}\Lambda(T_{\rm e,down}(t))}\approx5.0\times10^5\left(\frac{k_B T_{\rm e,down}(t)}{1.0\, {\rm eV}}\right)\left(\frac{n_{\rm ion}\Lambda(T_{\rm e,down}(t))}{3.6\times10^{-18}~{\rm erg\,s^{-1}}}\right)^{-1}~\sec,}
\end{equation}
respectively. 
Here, Eq. (\ref{eq:t_eq}) gives a timescale for the temperature equilibrium between electrons and ions~\citep[e.g.,][]{1956pfig.book.....S}, with $n_\mathrm{e}$ being the electron number density in the cgs unit, $T_\mathrm{ion,down}$ and $T_\mathrm{e,down}$ being the ion and electron temperature in K. 
We assume that the shocked SN ejecta {are} dominated by oxygen ($A_\mathrm{ion} = 16$ and $A_\mathrm{e} = 1/1840$).
A singly ionized state is inferred from the electron temperature at the immediate downstream, thus we set $n_\mathrm{e} = 4\rho_\mathrm{ej}(r_\mathrm{sh})/16m_\mathrm{u}$, $Z_\mathrm{ion} = 1$ and $Z_\mathrm{e} = 1$.  
While for the cooling timescale (Eq. \ref{eq:t_cool}), we employ the cooling function $\Lambda (T_\mathrm{e})$ for the oxygen-dominated gas given in \cite{2019MNRAS.489.4465K}~{(Figure \ref{fig:cooling_func})}.

\begin{figure}[ht]
    \centering
    \includegraphics[width=0.8\columnwidth]{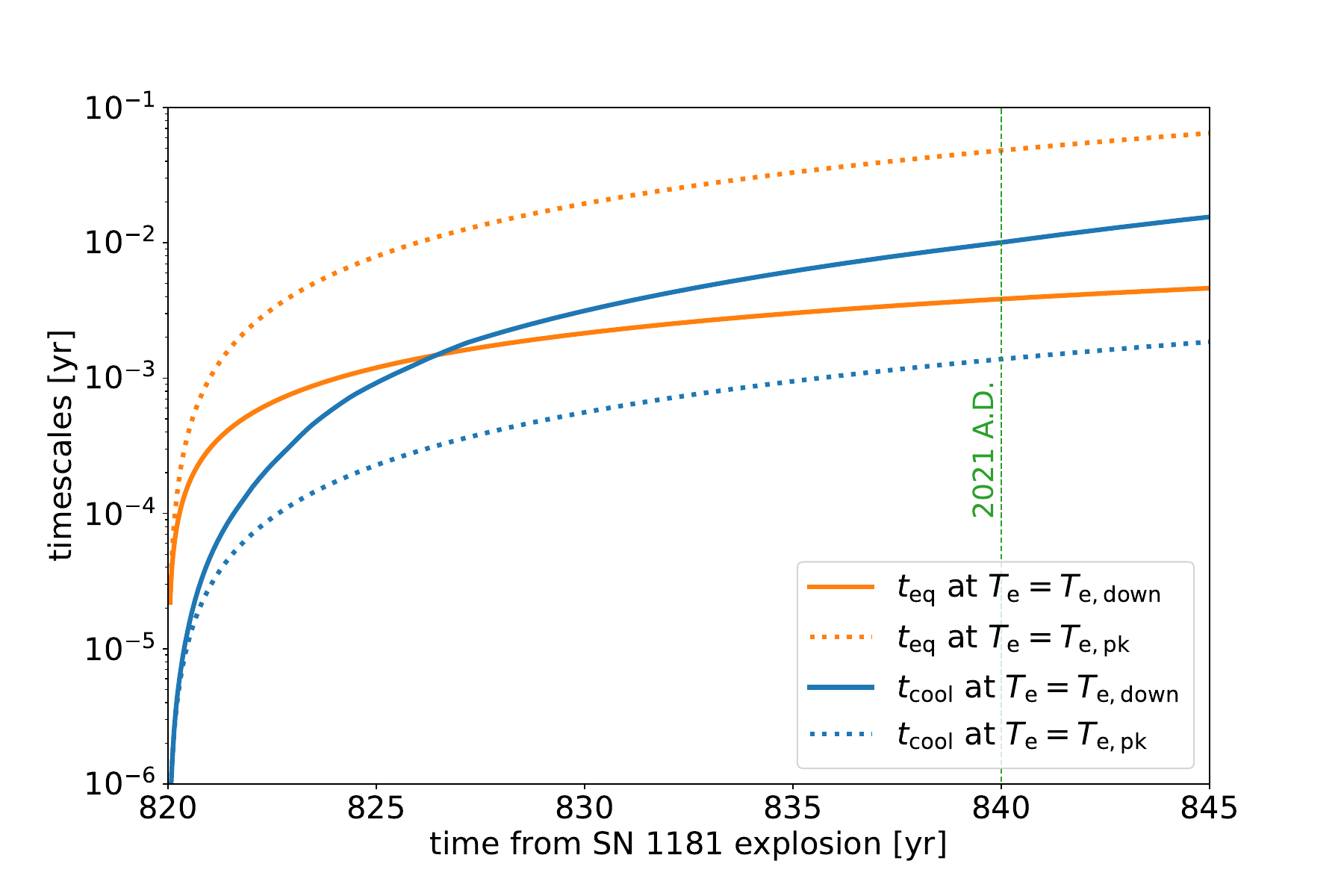}
    \caption{The time evolution of the heating (red) and cooling (blue) timescales of the inner shocked ejecta, calculated for a parameter set $(\dot{M}_{\rm w},t_{\rm w},M_{\rm{ej}}) = (1\times10^{-6}~M_\odot\,\yr^{-1},820~\yr,0.3~M_\odot)$ using our dynamical model. The solid lines are for the immediate shock downstream with an electron temperature of $T_\mathrm{e} = T_\mathrm{e, down}$ and the dotted lines are estimated for a peak temperature of the cooling function $k_\mathrm{B}T_\mathrm{e} = k_\mathrm{B}T_\mathrm{e, pk} = 3.7\,\mathrm{eV}$.}
    \label{fig:timescale}
\end{figure}

The solid lines in Figure \ref{fig:timescale} show the evolution of the heating and cooling timescales of electrons at the immediate shock downstream calculated for a parameter set $(\dot{M}_{\rm w},t_{\rm w},M_{\rm{ej}}) = (1\times10^{-6}~M_\odot\,\yr^{-1},820~\yr,0.3~M_\odot)$. 
{It can be observed that $t_\mathrm{eq}(T_\mathrm{e,down}) < t_\mathrm{cool}(T_\mathrm{e, down})$ at $t \sim 840\,\mathrm{yr}$. This implies that electrons at the immediate shock downstream with a temperature $T_\mathrm{e,down}$ will gradually approach equilibrium with ions, meaning that the electron temperature begins to increase due to collisional heating with ions.}

{On the other hand, the dotted lines in Figure \ref{fig:timescale} are the cooling and heating timescales of electrons at $T_\mathrm{e} = T_\mathrm{e,pk}$. We calculated these values in the same way as the equations (\ref{eq:t_eq}) and (\ref{eq:t_cool}).}
We find that the electron temperature is not likely to get into the X-ray regime, {i.e., $k_\mathrm{B}T_\mathrm{e,down} \ll 0.1\,\mathrm{keV}$}; 
the radiative cooling becomes relevant as the cooling function $\Lambda(T_\mathrm{e})$ steeply increases with electron temperature for $k_\mathrm{B}T_\mathrm{e, down}\sim 1\,\mathrm{eV} < k_\mathrm{B}T_\mathrm{e} < 3.7\,\mathrm{eV}$. 
It is shown that $t_\mathrm{eq}(T_\mathrm{e,pk}) > t_\mathrm{cool}(T_\mathrm{e, pk})$ is always satisfied, which means that the collisional heating of the electron saturates before reaching $k_\mathrm{B}T_\mathrm{e,pk} = 3.7\,\mathrm{eV}$. After the electron temperature becomes saturated at $T_\mathrm{e, down} < T_\mathrm{e} < T_\mathrm{e, pk}$, the ion temperature starts to decrease from $T_\mathrm{ion, down}$ via the radiative cooling. Since these heating and cooling timescales of the plasma are much shorter than the dynamical timescale of the shock $\approx r_\mathrm{sh}/v_\mathrm{sh} \approx \Delta t_\mathrm{w} =$ a few 10 yr, the shock downstream is radiative, i.e., the internal energy produced by the shock is predominantly lost via thermal radiation with a temperature of $T_\mathrm{e, down} < T < T_\mathrm{e, pk}$. 

\begin{figure}[ht]
    \centering
    \includegraphics[width=0.8\columnwidth]{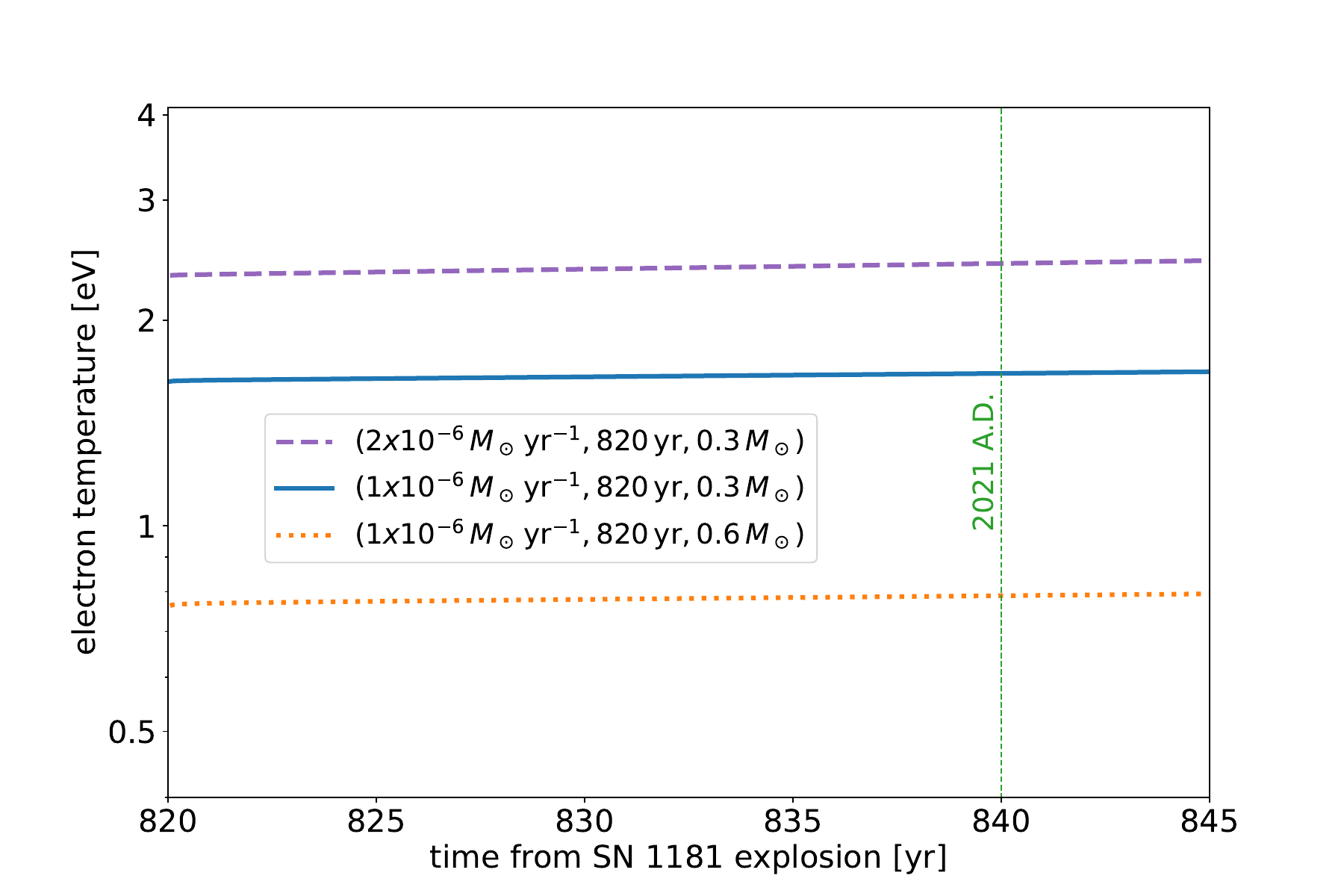}
    \caption{The time evolution of the {}{electron} temperature of the inner shocked SN ejecta obtained by the radiative shock condition (Eq. \ref{eq:rad_eq}).}
    \label{fig:electron_temperature}
\end{figure}

Although a radiative transfer calculation including the photo- and collisional ionization of the downstream plasma is required to accurately calculate the {}{electron} temperature, it can be roughly estimated in the radiative shock limit from the following condition; 
\begin{equation}\label{eq:rad_eq}
   \EM_{\rm ej}\Lambda(T_{\rm e}) \approx 2\pi r_{\rm sh}^2 \rho_{\rm ej}(r_\mathrm{ej})v_{\rm sh}^3,
\end{equation}
where 
\begin{equation}\label{eqn:EM_in_ej}
    \EM_{\rm ej} = n_{\rm e} n_{\rm ion} V_{\rm sh,ej}  = \frac{\rho_{\rm ej}^2V_{\rm sh,ej}}{\mu_{\rm e}\mu_{\rm ion}m_{\rm u}^2},
\end{equation}
is the emission measure of the shocked ejecta with 
\begin{equation}\label{eqn:ejecta_volume}
    V_\mathrm{sh,ej} \approx \frac{4}{3}\pi\left(r_\mathrm{in,f}^3-r_\mathrm{sh}^3\right)=\frac{4}{3}\pi r_\mathrm{sh}^3\left[\left(\frac{2\gamma}{1+\gamma}\right)^3-1\right]
\end{equation}
being its volume. The left and right hand sides of Eq. (\ref{eq:rad_eq}) represent the emission luminosity of and the energy injection rate into the shocked ejecta. 

Figure \ref{fig:electron_temperature} shows the {electron} temperature obtained from Eq. (\ref{eq:rad_eq}) for several sets of model parameters:   
{To estimate the electron temperature $T_\mathrm{e}$ from Eqs. (\ref{eq:rad_eq})-(\ref{eqn:ejecta_volume}), we need to begin by solving the shock dynamics using the model outlined in Section 3, which is characterized by model parameters $M_\mathrm{ej}$, $\dot M_\mathrm{w}$, and $t_\mathrm{w}$.}
As expected from the timescale arguments above, the emission temperature is in the range of $\lesssim 3\mbox{--}4\,\mathrm{eV}$, suggesting that the inner shocked SN ejecta emit mainly UV photons, not X rays. These UV photons are expected to be absorbed by the surrounding dust-rich ring and re-emitted in the IR bands. 

\bibliography{ref}{}
\bibliographystyle{aasjournal}
\end{document}